\documentclass[12pt]{article}
\usepackage[top=1in, left = 1in, bottom = 1in, right = 1in, paperheight = 11in, paperwidth = 8.5in]{geometry}

\usepackage[utf8]{inputenc}
\usepackage{adjustbox}
\usepackage{sectsty}
\usepackage[hyphens]{url}
\usepackage[breaklinks]{hyperref}
\usepackage{amsmath, amssymb, amsthm, amsbsy, amsfonts, cases}
\usepackage{graphicx, comment}
\usepackage[noend]{algpseudocode}
\usepackage{bigints}
\usepackage[mathscr]{euscript}
\usepackage[makeroom]{cancel}
\usepackage{graphicx}
\usepackage[table]{xcolor}
\usepackage{subcaption}
\usepackage{enumitem}
\usepackage{verbatim}
\usepackage[font={small,it,singlespacing}]{caption}
\usepackage{lscape}
\usepackage{caption}
\usepackage{hyperref}
\usepackage[sort]{natbib}
\usepackage{parskip}
\usepackage{array}
\usepackage{multirow}
\usepackage{graphicx}
\usepackage{wrapfig}
\usepackage{lscape}
\usepackage{rotating}
\usepackage{epstopdf}
\usepackage[compact]{titlesec}
\usepackage{authblk}
\usepackage[toc,page]{appendix}
\usepackage{setspace}
\usepackage{xr}
\usepackage[sort]{natbib}
\usepackage{color, colortbl}
\definecolor{Gray}{gray}{0.9}
\usepackage{algorithm}
\usepackage{algpseudocode}
\usepackage{boldline}
\usepackage[letterspace=-9]{microtype}

\hypersetup{colorlinks,citecolor=blue}
\graphicspath{{./images//}}

\setlist{nosep}

\sectionfont{\fontsize{12}{12}\selectfont}
\subsectionfont{\fontsize{12}{12}\selectfont}
\subsubsectionfont{\fontsize{12}{12}\selectfont}

\newcommand{\M}{\mathcal{M}}

\newcommand{\e}{\mathbf{e}}

\newcolumntype{L}[1]{>{\raggedright\let\newline\\\arraybackslash\hspace{0pt}}m{#1}}
\newcolumntype{C}[1]{>{\centering\let\newline\\\arraybackslash\hspace{0pt}}m{#1}}
\newcolumntype{R}[1]{>{\raggedleft\let\newline\\\arraybackslash\hspace{0pt}}m{#1}}

\theoremstyle{plain}
\newtheorem{thm}{Theorem}
\newtheoremstyle{exampstyle}
  {\topsep} 
  {\topsep} 
  {} 
  {} 
  {\bfseries} 
  {} 
  {.5em} 
  {} 
 
\newtheorem{defn}{Definition}
\newtheoremstyle{exampstyle}
  {\topsep} 
  {\topsep} 
  {} 
  {} 
  {\bfseries} 
  {.} 
  {.5em} 
  {} 

\providecommand{\keywords}[1]
{
  \small	
  \textbf{\textit{Keywords---}} #1
}

\usepackage{xr}
\makeatletter
\newcommand*{\addFileDependency}[1]{
  \typeout{(#1)}
  \@addtofilelist{#1}
  \IfFileExists{#1}{}{\typeout{No file #1.}}
}
\makeatother

\newcommand*{\myexternaldocument}[1]{
    \externaldocument{#1}
    \addFileDependency{#1.tex}
    \addFileDependency{#1.aux}
}
\myexternaldocument{supplemental}

\usepackage{algpseudocode}
\usepackage{algorithm}

\algnewcommand\algorithmicinput{\textbf{Input:}}
\algnewcommand\Input{\item[\algorithmicinput]}

\algnewcommand{\algorithmicoutput}{\textbf{return:}}
\algnewcommand\Output{\item[\algorithmicoutput]}

\algnewcommand\algorithmicforeach{\textbf{for each}}
\algdef{S}[FOR]{ForEach}[1]{\algorithmicforeach\ #1\ \algorithmicdo}

\theoremstyle{plain}

\title{\vspace{-0.5in}\normalsize{\textbf{A Feasibility Study of Differentially Private Summary Statistics and Regression Analyses with Evaluations on Administrative and Survey Data}}\vspace{-1em}}

\author[1]{\small{Andr\'es F. Barrientos}
}
\author[2]{\small{Aaron R. Williams}}
\author[3]{\small{Joshua Snoke}}
\author[2]{\small{Claire McKay Bowen}\vspace{-1em}}

\affil[1]{Florida State University\\ abarrientos@fsu.edu}
\affil[2]{Urban Institute\\ awilliams@urban.org \& cbowen@urban.org}
\affil[3]{RAND Corporation\\ jsnoke@rand.org}

\date{}

\begin{document}
\doublespacing
\lsstyle

\maketitle

\vspace{-30pt}

\noindent\textbf{Abstract:} Federal administrative data, such as tax data, are invaluable for research, but because of privacy concerns, access to these data is typically limited to select agencies and a few individuals. An alternative to sharing microlevel data is to allow individuals to query statistics without directly accessing the confidential data. This paper studies the feasibility of using differentially private (DP) methods to make certain queries while preserving privacy. We also include new methodological adaptations to existing DP regression methods for using new data types and returning standard error estimates. We define feasibility as the impact of DP methods on analyses for making public policy decisions and the queries accuracy according to several utility metrics. We evaluate the methods using Internal Revenue Service data and public-use Current Population Survey data and identify how specific data features might challenge some of these methods. Our findings show that DP methods are feasible for simple, univariate statistics but struggle to produce accurate regression estimates and confidence intervals.
To the best of our knowledge, this is the first comprehensive statistical study of DP regression methodology on real, complex datasets, and the findings have significant implications for the direction of a growing research field and public policy.

\section{Introduction}\label{sec:intro}
Federal tax data, derived from individuals' and businesses' tax and information returns, are invaluable resources for understanding the economic effects of policies, social and economic forces, and more on individuals, families, and firms. For example, \citet{chetty2014measuring} used tax data to study economic mobility across generations and how elementary school teacher quality affects economic outcomes later in life. However, full access to these data is available only to select government agencies, to a very limited number of researchers working in collaboration with analysts in those agencies, or through highly selective programs within the Internal Revenue Service (IRS) Statistics of Income (SOI) Division. Additionally, the manual process of vetting each statistical release for disclosure risks is labor intensive and imperfect because it relies on subjective human review.

Typically, researchers access federal confidential data in two ways: (1) direct access to the confidential data or (2) public statistics and public data files. This framework for accessing data is normal for most federal statistical agencies in the United States, such as the U.S. Census Bureau and Bureau of Labor Statistics. However, both options can have major drawbacks. For example with tax data, researchers at SOI must undergo an extensive clearance process to gain direct access to the confidential data. However, this clearance process can take several months and has eligibility requirements, such as being a U.S. citizen. Even after being cleared, researchers must access the data through a secure research enclave, such as the Federal Statistical Research Data Centers, which can be located hundreds of miles away.\footnote{One of the authors is 417 miles away from the closest Federal Statistical Research Data Center.} When releasing public statistics and public data files, SOI has progressively restricted and distorted the public information over the years due to general concern for protecting participants' data privacy \citep{bryant2014design}. 

\subsection{Practical Relevance}\label{subsec:bkg}
Expanding privacy-preserving access to confidential administrative data beyond these traditional means will be vital for scientific progress. For instance, \cite{nagaraj2023does} studied the impact of confidential administrative data on the rate, direction, and policy relevance of economics research. One of their findings showed that applied researchers with confidential data access through the Federal Statistical Research Data Centers produced 24\% more publications in top journals per year. Their work also highlighted that researchers with confidential administrative data access are more likely to produce papers which receive more citations in public policy documents.

The concept of validation servers provides a new tier of administrative data access. 
A validation server is a system that allows users to submit and run statistical queries on the confidential (validation) data after the users have developed their queries using theory or public data.\footnote{Not to be confused with verification servers which allow users to verify public data estimates against the confidential data without return confidential estimates. For example, \cite{barrientos2018providing} created a pilot verification server for the U.S. Office of Personnel Management} As an example of a validation server, until September 2022, the U.S. Census Bureau provided a framework to validate analyses for two experimental synthetic databases via the Synthetic Data Server at Cornell University: the Synthetic Longitudinal Business Database\footnote{``Synthetic Longitudinal Business Database (SynLBD),'' https://www.census.gov/programs-surveys/ces/data/public-use-data/synthetic-longitudinal-business-database.html} and the Survey of Income and Program Participation's (SIPP) Synthetic Beta Data Product\footnote{``Synthetic SIPP Data,'' https://www.census.gov/programs-surveys/sipp/guidance/sipp-synthetic-beta-data-product.html} \citep{benedetto2013creation, drechsler2014synthetic}.

A validation framework would allow researchers to access otherwise restricted data. The backbone of a validation server is statistical queries, such as univariate means or coefficients and confidence intervals for regression analyses. To protect privacy, agency staff must still review the queries for disclosure risks, which can delay releasing the results to the researchers. To address the manual review issue, one might adopt methods that satisfy differential privacy (DP) \citep{dwork2006calibrating}, which enables privacy-loss accounting and permits cumulative confidentiality risk control. Because DP privacy-loss guarantees do not depend on the confidential data and because the DP frameworks discussed in this paper all compose, a well-constructed validation server that implemented DP mechanisms consistent with our definitions could use the privacy-loss accounting enabled by composition to replace manual review. 
As of yet no work has considered the practical feasibility of current DP methods for querying summary statistics or regression analyses, which is the motivation for this paper.

\subsection{Contributions from this Paper}\label{subsec:cont}
We conduct an extensive study on state-of-the-art DP mechanisms to understand the feasibility of current DP methods for querying summary statistics and regression analyses. Our choice of queries is motivated by the use case of a validation server for tax policy research. Accordingly, we test methods for tabular statistics, mean statistics, quantile statistics, and statistics from regression analyses with cross-sectional data. 
There are several other analyses we have identified that we did not test, such as model selection, regression discontinuity, and kink designs. These methods are important for tax policy researchers, but the current DP methodology for these techniques is either in its early stages of development or does not support them.

We test the DP methods on the SOI Public Use File (SOI PUF) and Current Population Survey to measure feasibility. Importantly, our findings largely hold on both data sets. The SOI PUF is an annual file of sampled individual income tax returns after applying privacy protections. Several organizations, such as the American Enterprise Institute, the Urban-Brookings Tax Policy Center, and the National Bureau of Economic Research develop SOI PUF-based microsimulation models that help inform the public on potential impacts of policy proposals \citep{mcclelland2019tcja, debacker2019integrating, bierbrauer2021politically}. Because access to the public file is limited, we cannot provide the full data for others to replicate our study results using the SOI PUF. For this reason, the results presented in the main body of this paper use the 1994 to 1996 Current Population Survey Annual Social and Economic Supplements (CPS ASEC), which are publicly accessible through IPUMS USA \citep{ruggles2021cps}. The CPS ASEC has similar variables as the SOI PUF, allowing for similar queries to those we run on the SOI PUF. We provide parallel results using the SOI PUF in the Supplemental Materials. Additionally, we have a public repository containing all of our code, data, and results.\footnote{GitHub repo website, https://github.com/UrbanInstitute/formal-privacy-comp-appendix}


We develop selection criteria that excludes methods which work in theory only under specific conditions that would not normally be met in practice or cannot be tested. We find that many of the proposed methods in the literature could not be practically used in our setting. When selecting DP algorithms, we provide descriptions of each method, consider their ease of implementation, determine whether they require any additional tuning parameters, and assess their computational feasibility. This assessment will be useful for the statistical data privacy community in developing more applicable DP methods for the future.

We also contribute new methodology for obtaining estimates using DP regression. Specifically, we improve the sensitivity calculation for fitting the models with binary or categorical predictors and provide means of obtaining standard error estimates in addition to point estimates. Many of the existing methods for DP regression only provide point estimates, but we require that the validation server supports statistical inference in addition to estimation. Further, obtaining standard errors or confidence intervals is crucial to most statistical analyses. 

We define feasibility based on the impact of DP methods on analyses for making public policy decisions and their accuracy according to several utility metrics. Though motivated by an application to administrative tax data, the results of this paper are useful for the general study of DP and the feasibility of DP queries. We evaluate the methods using real data and identify how specific data features might challenge some of these methods. To the best of our knowledge, this paper provides the first comprehensive case study on DP queries. DP is a rapidly growing and popular field of study, but the vast majority of the work has focused on theoretical developments. Our results and conclusions will be vital in informing the current state of DP methods, addressing practical problems, and identifying directions for future work, particularly for statistical inference. 

We organize the remainder of the paper as follows. Section \ref{sec:dp} provides the definitions and theorems from the relevant DP frameworks and the implementing mechanisms used in this paper. Section \ref{sec:DPalgorithforTax} examines several categories of DP methods, providing a thorough discussion on which methods met our selection criteria, the methodological extensions we provide, and which methods could not be implemented in practice. Section \ref{sec:case-study} discusses the case studies and assumptions we made about the data and data analysts' knowledge. Section \ref{sec:utility} covers how we define the data utility and what metrics we use to evaluate the DP methods. Sections \ref{sec:summary-results} and \ref{sec:regresion-results} compare the selected DP methods on the CPS ASEC data and determines the feasibility of using these methods for a validation server. Concluding remarks and areas for future work are given in Section \ref{sec:conclusion}. The Supplementary Materials contain additional technical details on the methods and expanded results for our case studies.

\section{Differential Privacy}\label{sec:dp}
Differential privacy (DP) is a mathematical framework for providing a provable and quantifiable amount of privacy protection. Various definitions of DP exist, and given a particular definition, one can prove whether a method satisfies that type of DP. We refer to such methods\footnote{A note on terminology: In the context of DP, the terms \textit{mechanism}, \textit{algorithm}, and \textit{method} are often used interchangeably to describe the process of releasing a private statistical output. 
We do not see a clear delineation in the literature when using the three terms. More crucially is that anything referred to as a DP method, DP mechanism, or DP algorithm must provably satisfy the relevant definition of DP.} as differentially private (DP\footnote{Note that we use DP as an acronym for both ``differential privacy" and ``differentially private".}) methods. Satisfying DP is a provable feature of a method, not the data---a common misconception.

In this section, we reproduce the different definitions of DP used in this study, and three key theorems that are used in many methods studied in this paper. We use the following notation: $X\in\mathbb{R}^{n\times r}$ is the confidential dataset representing $n$ data points and $r$ variables and $M:\mathbb{R}^{n\times r}\rightarrow\mathbb{R}^k$ denotes the statistical query, i.e.,  $M$ is a function mapping $X$ to $k$ real numbers. 
We denote randomized versions of $M$ as $\M$ with the same domain and range. When appropriately parameterized, randomized mechanisms $\M$ can be said to implement particular DP frameworks.

\subsection{Definitions of Differential Privacy}\label{subsec:def}
\begin{defn}\label{def:dp}
    \textbf{Differential Privacy} \citep{dwork2006calibrating}:
    A sanitization algorithm, $\M$, satisfies $\epsilon$-DP if for all subsets $S\subseteq Range(\M)$ and for all $X,X'$ such that $d(X,X')=1$, 
        \begin{equation}\label{eqn:dp}
            \frac{\Pr(\M( X) \in S)}{ \Pr(\M( X')\in S)}\le \exp(\epsilon)
        \end{equation}
    \noindent where $\epsilon>0$ is the privacy loss budget and $d(X,X')=1$ represents the possible ways that $X'$ differs from $X$ by one record.
\end{defn}

Definition \ref{def:dp} is known as $\epsilon$-DP. At a high level, DP links the potential for privacy loss to how much the answer of a query (e.g., statistic) is changed given the presence or absence of any possible person's data from any possible data set. The role of $\epsilon$ is to control the privacy loss. Intuitively, when $\epsilon$ decreases, the maximum distance between the probability distributions of $\M( X)$ and $\M( X')$ become smaller, indicating that $\M( X)$ and $\M( X')$  are less distinguishable in distribution. Hence, users cannot determine whether the mechanism's outputs are based on $X$ or $X'$, which in turn protects the confidential information of that record that distinguishes $X$ and $X'$. Thus, low values of $\epsilon$ indicate high privacy levels and vice versa. $\epsilon$-DP can also be interpreted from a more statistical perspective in the context of hypothesis testing and under both frequentist \citep{wasserman2010statistical} and Bayesian \citep{kasiviswanathan2014semantics} paradigms.

Two definitions exist on what it means to differ by one record \citep{kifer2011no}. One definition assumes the presence or absence of a record, where the dimensions of $X$ and $X'$ differ by one row, making $X$ and $X'$ unbounded neighbors. The other definition assumes a change in the value of one record, where $X$ and $X'$ have the same dimensions, making $X$ and $X'$ bounded neighbors. \citet{kifer2011no} refers to these as \textit{unbounded DP} for presence or absence of a record and \textit{bounded DP} for the change of a record. \citet{li2016differential} state that unbounded DP satisfies an important composition theorem, which we will discuss later in this section (see Theorem \ref{thm:comp}), whereas bounded DP does not. In this paper, we assume unbounded DP, because we rely on Theorem \ref{thm:comp}.

Several relaxations of $\epsilon$-DP have been developed in order to inject less noise into the output, such as $(\epsilon, \delta)$-DP \citep{dwork2006our}, probabilistic DP \citep{machanavajjhala2008privacy}, concentrated DP \citep{dwork2016concentrated}, R\'enyi DP \citep{mironov2017renyi}, and zero-concentrated DP \citep{bun2016concentrated}. Although these definitions use the same provable privacy framework, they utilize alternative parameters offering different privacy guarantees. In return, they allow more possibilities for the type of noise added. 

\begin{defn}\label{def:adp} \textbf{$(\epsilon, \delta)$-Differential Privacy} \citep{dwork2006our}:
A sanitization algorithm, $\M$, satisfies $(\epsilon, \delta)$-DP if for all $X, X'$ that are $d(X,X')=1$,
    \begin{equation}\label{eqn:adp}
        \Pr(\M( X) \in S)\le \exp(\epsilon) \Pr(\M( X')\in S) + \delta
    \end{equation}
    where $\delta\in [0,1]$. 
\end{defn}

Definition \ref{def:adp} provides a simple relaxation of Definition \ref{def:dp} by adding the parameter $\delta$, which allows the privacy loss associated with the $\epsilon$ bound to fail at a rate no greater than $\delta$. Definition \ref{def:dp} can also be defined as a special case of $(\epsilon, \delta)$-DP when $\delta=0$.

\cite{dwork2016concentrated} proposed concentrated DP, which aimed to reduce the total privacy loss over multiple queries. \cite{bun2016concentrated} later improved this definition of privacy with zero-concentrated DP (zCDP or $\rho$-zCDP), given in Definition \ref{def:zcdp}. 

\begin{defn}\label{def:zcdp} \textbf{Zero-Concentrated Differential Privacy} \citep{bun2016concentrated}:
A sanitization algorithm, $\M$, satisfies $(\xi, \rho)$-zero-concentrated differential privacy if for all $X, X'$ that are $d(X,X')=1$ and $\alpha\in (1, \infty)$,
    \begin{equation}\label{eqn:zcdp}
        D_\alpha(\M(X)||\M(X'))\leq\xi+\rho\alpha,
    \end{equation}
    where $D_\alpha(\M(X)||\M(X'))$ is the $\alpha$-R\'enyi divergence  between the distribution of $\M(X)$ and $\M(X')$. $\rho$-zCDP is a special case of $(\xi, \rho)$-zCDP when $\xi=0$.
\end{defn}

\cite{bun2016concentrated} provide results that allows us to relate $\epsilon$-DP and $(\epsilon, \delta)$-DP algorithms to $\rho$-zCDP equivalents. They show in their Proposition 4 that if $\M$ satisfies $\epsilon$-DP, then $\M$ satisfies $(\frac{1}{2}\epsilon^2)$-zCDP. They also show in their Proposition 3 that if $\M$ satisfies $\rho$-zCDP, then $\M$ is $(\rho+2\sqrt{\rho\log(1/\delta)},\delta)$-DP for any $\delta>0$. Notice that, while $\rho$-zCDP also satisfies the definition of the $(\epsilon, \delta)$-DP, $\rho$-zCDP are more general. Every value of $\rho$ is associated with a continuum of $(\epsilon, \delta)$ values, which summarize the behavior of implementing mechanism. In their Lemma 3.6, they present a slight strengthening of this result stating that if $\M$ satisfies $(\xi,\rho)$-zCDP, then $\M$ is $(\xi + \rho+\sqrt{4\rho\log(\sqrt{\pi \rho}/\delta)},\delta)$-DP for any $\delta>0$. We use the result of this Lemma to express any $\rho$-zCDP algorithm in terms of $(\epsilon, \delta)$-DP. For a given  $(\epsilon, \delta)$, we must find the value of $\rho$ that satisfies $\epsilon = \rho+\sqrt{4\rho\log(\sqrt{\pi \rho}/\delta)}$.

\subsection{Global Sensitivity and Differentially Private Mechanisms}\label{subsec:mech}
In this section, we introduce the concept of global sensitivity and present three of the fundamental mechanisms that satisfy $\epsilon$-DP and $(\epsilon, \delta)$-DP and form the building blocks of many of the DP algorithms we test in this paper.

Independent of the values of $\epsilon$ and $\delta$, an algorithm that satisfies $\epsilon$-DP or $(\epsilon, \delta)$-DP must adjust the amount of noise added to the output based on the maximum possible change between any two databases that differ by one row. This is commonly referred to as the global sensitivity (GS), given in Definition \ref{def:gs}.

\begin{defn}\label{def:gs} \textbf{$l_1$-Global Sensitivity} \citep{dwork2006calibrating}:
For all $X,X'$ such that $d(X,X')=1$, the global sensitivity of a function $M$ is
    \begin{equation}\label{eqn:gs}
        \Delta_1 (M)= \underset{d(X,X')=1}{\text{sup}} \|M(X)-M(X') \|_1 
    \end{equation}
\end{defn}

We can calculate global sensitivity under different norms. For instance, $\Delta_2(M)$ represents the $l_2$ norm GS ($l_2$-GS) of the function $M$. 
Although the definition is straightforward, calculating the GS can often be difficult in practice. For instance, we cannot calculate a finite GS of one of the most common statistics, the sample mean, if the variable is not bounded (or the bound is not known).

A commonly used mechanism satisfying $\epsilon$-DP is the Laplace mechanism, given in Definition \ref{def:lap}. \cite{dwork2006calibrating} proved it satisfies $\epsilon$-DP and uses the $l_1$-GS. Another popular mechanism is the Gaussian mechanism, given in Definition \ref{def:gauss}, which uses the $l_2$-GS of the statistical query. \cite{dwork2014algorithmic} showed the Gaussian mechanism satisfies $(\epsilon,\delta)$-DP. 

\begin{defn}\textbf{Laplace mechanism} \citep{dwork2006calibrating}: \label{def:lap}
Given any function $M:\mathbb{R}^{n\times r}\rightarrow\mathbb{R}^k$, the Laplace mechanism is defined as:
    \begin{equation}\label{eqn:lap}
        \M(X)=M(X)+(\eta_1,\ldots,\eta_k).
    \end{equation}
where $(\eta_1,\ldots,\eta_k)$ are i.i.d. $Laplace(0, \frac{\Delta_1(M)}{\epsilon})$.
\end{defn}

\begin{defn}\label{def:gauss} \textbf{Gaussian mechanism} \citep{dwork2014algorithmic}:
Given any function $M:\mathbb{R}^{n\times r}\rightarrow\mathbb{R}^k$, the Gaussian mechanism is defined as:
    \begin{equation}\label{eqn:gauss}
        \M(X)=M(X)+(\eta_1,\ldots,\eta_k).
    \end{equation}
where $(\eta_1,\ldots,\eta_k)$ are i.i.d. $N\left(0, \sigma^2 = \left(\frac{\Delta_2(M)\sqrt{2 \log(1.25/\delta)}}{\epsilon}\right)^2\right)$.
\end{defn}

Both the Laplace and Gaussian mechanisms are simple and quick to implement, but they apply only to numerical values (without additional post-processing). A more general mechanism which satisfies $\epsilon$-DP is the Exponential mechanism, given in Definition \ref{def:exp}, which allows for the sampling of values from a noise-infused distribution rather than adding noise directly.

 \begin{defn}\label{def:exp} \textbf{Exponential mechanism} \citep{mcsherry2007mechanism}:
     The Exponential mechanism releases values with a probability proportional to
         \begin{equation}\label{eqn:exp}
             \exp \left(\frac{\epsilon u(X, \theta)}{2\Delta_1(u)}\right),
         \end{equation}
     where $u(X,\theta)$ is a quality function that determines the values for each possible output, $\theta$, on $X$.
 \end{defn}
 
\subsection{Composition and Post-Processing Theorems}\label{subsec:thm}
Lastly, we introduce the important concepts of composition and post-processing. These enable the development of more complex algorithms that combine DP mechanisms with post-processing in order to release multiple statistics with additional structural or noise-reducing enhancements
The composition theorems given in Theorem \ref{thm:comp} formalize the concept of totaling the privacy loss incurred across multiple queries or datasets.

\begin{thm}\label{thm:comp} \textbf{Composition Theorems} \citep{mcsherry2009privacy,dwork2016concentrated,bun2016concentrated}:
Suppose a mechanism, $\M_j$, provides $(\epsilon_j$, $\delta_j)$-DP or $(\xi_j,\rho_j)$-zCDP for $j=1,\ldots,J$.
  \begin{itemize}\setlength{\itemindent}{15pt}
  \item[a)] \textbf{Sequential Composition:} The sequence of $\M_j(X)$ applied on the same $X$ provides $(\sum_{j=1}^J\epsilon_j,\sum_{j=1}^J\delta_j)$-DP or $(\sum_{j=1}^J\xi_j,\sum_{j=1}^J\rho_j)$-zCDP.
  \item[b)] \textbf{Parallel Composition:} Let  $X_j$ be disjoint subsets of the dataset $X$, $j=1,\ldots,J$. The sequence of $\M_j(X_j)$ provides $(\max_{j \in \{1,\ldots,J\}} \epsilon_j, \max_{j \in \{1,\ldots,J\}} \delta_j)$-DP or \\ $(\max_{j \in \{1,\ldots,J\}}\xi_j, \max_{j \in \{1,\ldots,J\}}\rho_j)$-zCDP.
  \end{itemize}
\end{thm}

If we want to make $J$ statistical queries on $X$ and we want the total privacy loss to equal $\epsilon$, the composition theorems state under what conditions we may allocate portions of the overall $\epsilon$ to each statistic. Under sequential composition, a typical appropriation is dividing $\epsilon$ and $\delta$ equally by $J$. For example, a data practitioner might want to query the mean and standard deviation of a variable. These two queries will require using the sequential composition, allocating an equal amount of privacy budget to each query. \cite{dwork2010boosting} and \cite{bun2016concentrated} proposed other forms of sequential composition, but the methods we test do not rely on these works.

Conversely, parallel composition does not require splitting the budget because the noise is applied to disjoint subsets of the input domain. For example, privacy experts will often leverage parallel composition to sanitize histogram counts, assuming that the bins are disjoint subsets of the data. Noise can then be added to each bin independently without needing to split $\epsilon$ or $\delta$.

Theorem \ref{thm:post} (post-processing) states that any function applied to the output of a DP mechanism also satisfies DP. Some DP methods, as will be shown later, use the post-processing theorem to correct any inconsistencies or values that are not possible and to compute additional summaries required to perform statistical inference.

\begin{thm}\label{thm:post} \textbf{Post-Processing Theorem} \citep{dwork2006calibrating,nissim2007smooth, bun2016concentrated}:
If $\M$ be a mechanism that satisfies $(\epsilon,\delta)$-DP or $(\xi,\rho)$-zCDP, and $g$ be any function, then $g\left(\M(X)\right)$ also satisfies $(\epsilon,\delta)$-DP or $(\xi,\rho)$-zCDP.
\end{thm}

\subsection{Setting the Privacy Loss Budget}
When considering a validation server that releases results from numerous statistical queries, we must accurately account for the total privacy loss across all queries. To accomplish this, all mechanisms must use common privacy parameters. Accordingly, we show that every DP method considered in this paper satisfies $(\epsilon, \delta)$-DP. This means if we institute a validation server using the methods presented in this paper, the total privacy loss across all queries can be tracked by the total $\epsilon$ and $\delta$ used from each individual query. We include a longer discussion on the issues surrounding setting the privacy loss budget in the Supplementary Materials.

\section{Differentially Private Algorithms for a Tax Data Validation Server}\label{sec:DPalgorithforTax}
For this section, we review the DP methods that we implement for the feasibility study. We focus on methods that directly return the specific statistic. We do not consider alternative paradigms, such as creating bespoke DP histograms to calculate statistics \citep{foote2019releasing}. However, a validation server with sufficient computational capabilities could consider this in future work.

\subsection{Tabular Statistics}\label{subsec:tab}
The literature has shown that adding Laplace noise produces the most accurate estimates for single tabular queries. For example, \cite{rinott2018confidentiality}, \cite{bowen2021differentially}, \cite{shlomo2018statistical}, and \cite{liu2018generalized} found the Laplace mechanism is still ``hard to beat'' for disseminating frequency tables when the data have a large number of observations, or there are a lot of parameters and attributes. One can query the counts for a histogram jointly or as separate counts. We explore both because under separate queries \citet{mironov2017renyi} showed composition rules that reduce the overall variability of the Gaussian mechanism if the $l_2$-GS is 1 (which is true in the counts we consider). Ultimately, we found that joint queries outperformed separate queries, even with the advanced composition.\footnote{These results can be replicated using the code on GitHub at https://github.com/UrbanInstitute/formal-privacy-comp-appendix}
Therefore, we only present results for joint tabular statistics using the Laplace and Gaussian mechanisms.

\subsection{Quantile Statistics}\label{subsec:quant2}
The methods used for generating DP quantiles use either the Laplace mechanism or the Exponential mechanism. For instance, \cite{smith2011privacy} proposed an algorithm, \textit{IndExp}, for selecting individual quantiles using the Exponential mechanism that satisfies $\epsilon$-DP. \textit{IndExp} has since been implemented in the \cite{smartnoise} and \cite{ibm} DP libraries. \cite{gillenwater2021differentially} recently extended this method to two other algorithms, \textit{AppIndExp} and \textit{JointExp}. The former is the same as \textit{IndExp} but optionally satisfies $(\epsilon, \delta)$-DP using the composition theorem from \cite{dong2020optimal} to choose an optimal $\epsilon$ for multiple queries given a total ($\epsilon$, $\delta$). \textit{JointExp} samples multiple quantiles jointly, satisfying $\epsilon$-DP, and does not need composition for multiple quantiles. Using the Laplace mechanism, \citet{nissim2007smooth} developed an approach satisfying $(\epsilon, \delta)$-DP for sampling median values using smooth sensitivity that can be extended to query any other quantile. We will hereafter refer to this method as \textit{Smooth}. We do not test other proposed approaches from \citep{dwork2009differential,tzamos2020optimal} because they require fine-tuning based on distributional assumptions that a researcher might not realistically have in a validation server setting.

\subsection{Means and Confidence Intervals}\label{subsec:mean}

For mean estimates, we review DP methods that release means with their associated confidence intervals (CI). We find that common approaches in the literature use the Laplace mechanism, Gaussian mechanism, or Exponential mechanism for releasing some means with CIs. \cite{du2020differentially} conducted a comprehensive research study that aimed to move the theory to practice for releasing private CIs. The authors developed five new methods, two based on directly applying Laplace noise named, \textit{NOISYVAR} and \textit{NOISYMAD}, and three based on querying quantiles from the Exponential mechanism to estimate the standard deviation called, \textit{CENQ}, \textit{SYMQ}, and \textit{MOD}. \citet{du2020differentially} also compared their methods against methods developed by \citet{karwa2017finite}, \citet{d2015differential}, and \citet{brawner2018bootstrap}.

\citet{du2020differentially} proposed two other DP methods for releasing mean estimates uses quantiles, but the methods assume the data are very nearly Gaussian. We choose to not test these methods on our heavily skewed data because our preliminary tests show poor performance. Although transformations can be used to convert skewed data into approximate Gaussian form, they are not useful if the goal is to make inferences about the mean in the original scale, as assumed here. Other methods require more information than is realistic for our application. For instance, \cite{karwa2017finite}, \cite{bowen2020comparative}, \cite{d2015differential}, and \cite{biswas2020coinpress} require the researcher to set bounds on the standard deviation to calculate the GS. In a validation server setting, a user might not have any information on these bounds, so we do not test these methods for our study.

\subsection{Differentially Private Regression Analyses}\label{subsec:reg}
For this subsection, we begin with an overview of the current DP approaches for regression estimates. Specifically, we center our attention on the linear model $Y = X\beta+\e$, where $Y$ is the vector representing the observations for the response, $X$ is the design matrix, and $\e$ is the vector of independent and identically distributed normal errors. We then explain the criteria for including methods in this feasibility study, discuss the selected methods, and detail any adaptions required to include the methods in the experiments.

\subsubsection{Traditional Differentially Private Approaches for Regression Analyses}\label{subsec:reg_rev}

We classify DP methods for regression analysis according to the outputs they produce: (1) point estimates only, (2) point and interval estimates, and (3) other outputs related to regression analysis, such as diagnostic plots. Because we are particularly interested in methods that provide full statistical inference, we focus our study on methods from category (2) and discuss these methods in greater detail. We review the other two types further in the Supplemental Materials.

\citet{sheffet2017differentially} developed $(\epsilon,\delta)$-differentially private algorithms that, with certain probability, output summary statistics used for either traditional linear regression or ridge regression. When the outputs are summaries for ridge regression, the penalization parameter is a function of the algorithm's inputs instead of being predefined by the user. \citet{sheffet2017differentially} derived CIs and t-statistics that account for the noise added to the confidential summaries.
In addition, they included a second algorithm that adds Gaussian noise directly to the sufficient statistic $S=[X,Y]^t[X,Y]$ and showed that one could obtain CIs for the regression coefficients under certain conditions (i.e., the norm of the true regression coefficients is upper bounded). Despite this method's promise, the paper does not provide practical guidance on defining the additional tuning parameters that are essential for guaranteeing the correct confidence or significance level of the CIs and t-statistics. The lack of obtainable code is another reason we exclude it from our study.

\cite{sheffet2019old} proposed a $(\epsilon,\delta)$-DP mechanism that provides estimates with random noise drawn from the Wishart distribution. The mechanism defines a noisy statistic that preserves the property of being positive-definite, a common problem when adding noise to the regression sufficient statistics. According to the author, this mechanism is only valid for $\epsilon\in (0,1)$ and $\delta \in (0,\frac{1}{\exp(1)})$. 

\cite{wang2019differentially} also developed $\rho$-zCDP and $\epsilon$-DP methods that release noisy versions of the summary statistics while preserving positive definiteness. These methods add noise using either a normal distribution (for $\rho$-zCDP) or the spherical analogue of the Laplace distribution (for $\epsilon$-DP). The positive definiteness is achieved by using eigenvalue decomposition and censoring the eigenvalues falling below a given threshold. Although \cite{sheffet2019old} and \cite{wang2019differentially} did not derive CIs or t-values under the proposed mechanisms for the normal linear model, these contributions paved the way to develop DP methods that allow full inference for regression coefficients.

\cite{ferrando2020general} developed a general approach to produce point and interval estimates for different DP mechanisms applied to linear regression queries. The paper outlines two $\epsilon$-DP strategies that employ a noisy version of the sufficient statistics. The first approach applies the noisy statistics in classic ordinary least squares point and interval estimators. This strategy's accuracy and coverage is ensured by large-sample arguments. The second strategy also uses plug-in estimators, but computes CIs using a parametric bootstrap. The method accounts for both the injected noise and the underlying sampling distributions. 

A drawback of these two approaches is the lack of clarity on computing the points and interval estimates of the coefficients when the inverse of the noisy covariance matrix is not positive-definite. We adapt the approach from \cite{ferrando2020general} for this study by applying a regularized version of the noisy sufficient statistic to handle the positive definiteness issue. We provide details on how we use regularization in Section \ref{subsec:reg_meth}. We acknowledge that a new version of \cite{ferrando2020general} was recently released, where the authors made the update after we completed our feasibility study \citep{ferrando2021general}. Although there are slight differences between the main algorithms in each version, the first version of the algorithm produces valid results. We provide a full description of the algorithm in the Supplementary Materials.

For Bayesian inference, the developments in DP regression models are still at an early stage. 
We briefly describe two proposed approaches that can be used for full inference in the context of regression and discuss the reasons for excluding them from the feasibility study. First, \citet{bernstein2019differentially} offered an approach that relies on a noisy version of the sufficient statistics. The procedure uses a Bayesian framework and employs a large-sample distributional characterization of the sufficient statistics. Using a Bayesian framework allows the authors to draw from the regression coefficient's posterior distribution and, thus, provide point and interval estimates. This approach requires prior knowledge of the predictors' second and fourth crossed population moments. A researcher can overcome this issue by privately querying the corresponding sampling moments, eliciting a prior for the moments, or assigning a distribution to the predictors. Privately querying the sample moments will require an exponentially increasing portion of the total privacy budget, because the number of required moments grows exponentially with the number of predictors. This leads to exponentially increasing the noise with respect to the number of parameters in the model, assuming the privacy budget remains fixed. Also, as stated by the authors, the approach does not account for the noise added to the moments and may introduce some miscalibration of uncertainty. The other two alternatives require eliciting a prior for the moments or specifying a distribution for the predictor. Neither is generally feasible for all models, so this approach would have limited use in the validation server framework. For these reasons, we decided not to consider \citet{bernstein2019differentially}'s method in the feasibility study.


\citet{wang2018revisiting} also presented a method that involves drawing from the posterior distribution of the regression coefficients. The main concept of the method is centered around providing a single-point estimate of the coefficients using a single draw from the posterior distribution. However, it is tempting to use this approach for assessing uncertainty. Nevertheless, this poses a challenge since each draw from the posterior consumes a portion of the privacy budget. This limitation significantly hampers the practicality of \citet{wang2018revisiting}'s algorithm when it comes to assessing uncertainty. In practice, drawing thousands of values from the posterior to accurately capture uncertainty would necessitate dividing the privacy budget into thousands of small values. As a consequence, unless the analyst has an enormous privacy budget (which is unlikely in a validation server scenario), excessively splitting the privacy parameter will diminish the statistical utility of the method to a significant extent.



\subsubsection{Selected Methods and Adaptations}\label{subsec:reg_meth}

When selecting methods, we prioritize the feasibility of implementation before evaluating the statistical usefulness of the outputs. Specifically, our systematic approach to selecting the methods are as follows:

\noindent \textbf{1.} Check if the method described in the manuscript can be used for linear regression models with normal errors, can handle multiple predictors, and permit full inference on the estimates. Discard methods that, for example, only provide point estimates or only provide t-values for regression coefficients without corresponding point estimates.

\noindent \textbf{2.} Verify if the method achieves DP under testable assumptions. Discard methods that, for example, only achieve DP when the sample size is ``large enough," where the specific definition of ``large enough" is unknown, or under non-testable conditions related to the underlying data generating mechanism.

\noindent \textbf{3.} Determine if the manuscript provides all the required elements for implementating the algorithm. Discard methods whose manuscripts lack the necessary information needed to implement the proposed approach. This includes cases where pseudocode is absent and the implementation requires non-trivial choices that reflect the proposed method's underlying theory. Examples of non-trivial choices may involve decisions on selecting the dimension in a lower-dimensional projection or determining what to do when the statistics of interest take values on the cone of positive matrices, but the resulting noisy statistic falls outside of it after adding noise. In such cases, check if the required information is available in the description of the numerical experiments or illustrations within the manuscript. If unavailable, reach out to the authors to inquire about code implementing the method or for guidance on making such choices. If we do not receive a response, we then discard the method.

\noindent \textbf{4.} Identify if there are any issues in the algorithm implementing the method and, if possible, provide a potential solution to address them. Reach out to the authors to inquire whether our proposed solution is correct or if they can provide an alternative solution. If we do not receive a response, we then discard the method.

Based on this selection procedure, we assess each of the methods discussed in Section \ref{subsec:reg_rev}. Without additional adaptations, only one method, \citet{ferrando2020general}, meets all the inclusion criteria. 
We include \citet{brawner2018bootstrap}'s method, because, even though it was not originally designed for linear regression, a small adaptation makes it eligible. We also modify both \citet{ferrando2020general}'s and \citet{brawner2018bootstrap}'s methods (we refer to the latter as {BHM}) and obtain new versions of the methods to compare in the feasibility study. 

We repurpose elements of the algorithm from \cite{ferrando2020general} to perform full inference with other mechanisms, increasing the number of testable methods from two to six. \cite{ferrando2020general}'s approach employs a parametric bootstrap to approximate the distribution of the coefficients' estimator while accounting for the underlying data-generating distribution and the DP mechanism (see Algorithm 3 in \cite{ferrando2020general}). Although the original method uses the Laplace mechanism, we can employ this same technique with other DP mechanisms after making some simple adaptations. Specifically, we adapt Algorithm 3 to be compatible with the Analytic Gaussian mechanism from \cite{balle2018improving}, Algorithm 2 from \citet{sheffet2019old} (hereafter, the Wishart mechanism), and both approaches in Algorithm 2 from \cite{wang2019differentially} (hereafter, the Regularized Normal and Regularized Spherical Laplace mechanisms).

\cite{ferrando2020general}'s Algorithm 3 inputs are all functions of the noisy version of the summary statistics $S_H = S + H$, where $S=[X,Y]^t[X,Y]$ is the sufficient statistic for the linear model $Y = X\beta+\e$, $Y$ is the vector representing the observations for the response, $X$ is the design matrix, $\e$ is the vector of independent and identically distributed normal errors, and the elements of matrix $H$ are random noise sampled to achieve DP. The algorithm uses Monte Carlo simulation to approximate the sampling distribution of an estimator of $\beta$, say $\hat \beta_b$, based on $S_H$ that accounts for the randomness in both $\e$ and $H$. This algorithm assumes that the sample size $n$ is known, which may not be true in many applied scenarios. We choose to include the intercept in the linear model, so we obtain a privatized version of the sample size from the entry $(1,1)$ of $S_H$. We use this as the noisy sample size when applying \cite{ferrando2020general}'s Algorithm 3. If we chose not to include the intercept in our model, we would need to spend part of our privacy budget querying a noisy sample size estimate.

Implementing \cite{ferrando2020general}'s algorithm requires that $S_H$ be positive definite, which is not always guaranteed in practice. For this reason, we use regularization such that $S_H^r = S_H - rI_{(p+1)\times (p+1)}$, where $r$ is equal to zero if $S_H$ is positive definite or less than the minimum eigenvalue of $S_H$ otherwise. Thus, $S_H^r$ is guaranteed to be positive-definite. To find $r$ for the Wishart mechanism, we use Remark (2) of \cite{sheffet2019old}. The remark contains an analytic expression to ensure that $S_H^r$ is positive definite with probability greater than $1-\delta$. For the Laplace and Analytic Gaussian mechanisms implementations, we follow a similar idea and define $r$, such that $P[H - rI_{p\times p} \mbox{ definite positive}] \approx p_0$, where $p_0$ is close to one. For the Regularized Normal and Regularized Spherical Laplace mechanisms, we define $S_H^r$ as the regularized matrix resulting from \cite{wang2019differentially}'s Algorithm 2.

Unfortunately, we only have an analytical expression of $r$ for the Wishart mechanism. Hence, as proposed in \cite{pena2021differentially}, we employ simulations to approximate the distribution of the smallest eigenvalue of $H$ and define $r$ to be the $p_0$-percentile of this distribution. Note that $P[S_H^r \mbox{ definite positive}] > P[H- rI_{p\times p} \mbox{ definite positive}]$. When defining $r$ as before, $r$ sometimes fails to make $S_H^r$ positive definite, with small probability. Thus, we define $r$ as equal to three times the minimum eigenvalue of $S_H$, which we chose based on empirical testing. We provide our adaptation of \cite{ferrando2020general}'s algorithm in the Supplemental Materials.

Since the effect of $H$ on the inferences should diminish as the sample size increases, \cite{ferrando2020general} discusses another strategy to make inferences about $\beta$ that only accounts for the randomness in $\e$ but not in $H$. Under this strategy, the idea is to treat $S_H^r$ as if it was $S$. Let $\hat \beta$ represent the maximum likelihood estimator based $S$ (i.e., without DP), and let $\hat \beta_H$ denote an estimator of $\beta$ obtained by plugging-in $S_H^r$ in the formulas of the maximum likelihood estimator without DP. Theorem 1 in \cite{ferrando2020general} shows that $\hat \beta_H$ and $\hat \beta$ share the same asymptotic normal distribution and extends to mechanisms other than the Laplace mechanism, such as the Analytic Gaussian and Wishart. For this reason and to provide a method alternative to that based on $\hat \beta_b$ and its bootstrap-based distribution, we employ $\hat \beta_H$ and its asymptotic normal distribution to compute point and interval regression coefficient estimates. 

In addition to methods based on \cite{ferrando2020general}'s approach, we also implement the bootstrap approach described in Section 6.3 of \cite{brawner2018bootstrap}. We can directly apply this method when the summary statistic of interest is an average or a sum, such as $S$. Unfortunately, \cite{brawner2018bootstrap}'s strategy to compute CIs cannot be used in a straightforward manner for the CIs of regression coefficients.

However, since this method allows drawing multiple realizations of the noisy $S$ (after splitting the privacy parameters $(\delta, \epsilon)$), we can approximate the sampling distribution of the regression coefficients relying on an asymptotic assumption. Let $\hat \beta_{BH,\,1},\ldots,\hat \beta_{BH,\,K}$ represent $K$ sampled regression coefficients. Each of these coefficients is computed based on a realization of the noisy $S$ using the equivalent of $(\delta/K, \epsilon/K)$-DP under BHM approach. Thus, we approximate the sampling distribution of the regression coefficients by using a multivariate normal distribution, where the mean and covariance matrix correspond to the sample mean and covariance matrix based on $\hat \beta_{BH,\,1},\ldots,\hat \beta_{BH,\,K}$. We detail how we compute the sensitivity of $S$ for the Laplace mechanism in the Supplemental Materials.

In summary, we consider six methods for our feasibility study: the Laplace mechanism, the BHM, and adaptations to the Analytic Gaussian, Wishart, Regularized Normal, and Regularized Spherical Laplace mechanisms. Each method uses both of \cite{ferrando2020general}'s approaches except the BHM. We summarize the methods selected for evaluation in Table \ref{tab:reg}.

\begin{table}[!htp]
    \def\arraystretch{1}
    \caption{Summary of the differentially private regression methods we test for our tax use case studies. All methods return two types of confidence intervals (bootstrap-based and plug-in-based asymptotic) estimated using an adaptation from \citet{ferrando2020general}, with the exception of BHM.}
    \label{tab:reg}
    \centering
    \scriptsize
    \small
    \begin{tabular}{ C{5.5cm} | C{5.5cm} | C{4cm}}
        \hline
            \textbf{Method} & \textbf{Source} & \textbf{Privacy Definition}\\
        \hline
            Analytic Gaussian mechanism & \citet{balle2018improving} & $(\epsilon, \delta)$-DP \\
        \hline
            Laplace mechanism & \citet{ferrando2020general} & $\epsilon$-DP\\
        \hline
            Regularized Normal mechanism & Algorithm 2 \citep{wang2019differentially} & $\rho$-zCDP\\
        \hline
            Regularized Spherical Laplace mechanism & Algorithm 2 \citep{wang2019differentially} & $\epsilon$-DP\\
        \hline
            Wishart mechanism & Algorithm 2 \citep{sheffet2019old} & $(\epsilon, \delta)$-DP \\
        \hline
            Brawner-Honaker Method (BHM) & \citet{brawner2018bootstrap} & $\rho$-zCDP\\
        \hline
    \end{tabular}
\end{table}

\section{Data Used for Case Studies}\label{sec:case-study}
We evaluate the DP methods using two datasets. We use the 2012 SOI PUF to present results that will be most similar to a real validation server implementation using IRS data. SOI created this file based on a subsample of the confidential taxpayer data. The file contains 200 variables with 172,415 records and represents United States tax filers. The SOI PUF contains many variables with skewed distributions or have values predominantly zero, a notable feature for this study.

Because the SOI PUF is a restricted access file, we also test our DP algorithms on the 1994 to 1996 CPS ASEC data to ensure that other researchers can replicate the findings reported in this paper. The U.S. Census Bureau generates the CPS ASEC from a probability sample, which contains detailed information about income. Similarly to the SOI PUF, the CPS ASEC data has skewed variables and variables that are predominantly zeros. The CPS ASEC represents the U.S. civilian non-institutional population and has 91,500 households and 157,959 people. The biggest difference between the SOI PUF and CPS ASEC is that the latter reports information at the person level and household level instead of tax units. We do not expect this to have a large impact on our evaluations.

We present the CPS ASEC results in the main body of this paper and the SOI PUF results in the Supplemental Materials. Only results from our regression evaluations are presented here due to space constraints. We provide the results for other types of queries in the Supplementary Materials. The code to reproduce our results is available online.\footnote{GitHub repo website, https://github.com/UrbanInstitute/formal-privacy-comp-appendix}

\subsection{Tax Policy Case Studies}\label{sec:tax-study}
We base our tests on the types of analyses that tax economists would normally query for the SOI PUF data. For example, a tax expert would likely make one or more of the following queries:

\begin{itemize}
    \item \textbf{Counting query}: How many tax returns have salary and wage income in excess of \$100,000?
    \item \textbf{Mean statistic query}: What are the means for the total and subsets of the total population?
    \item \textbf{Quantile statistic query}: What is the income threshold for the top 10 percent of earners?
    \item \textbf{Regression query}: What is the relationship between education and annual earnings?
\end{itemize}

For regression analysis, we replicate a published cross-sectional multiple linear regression model from \cite{card1999causal}. We reproduce the models found from column (5) in Table 1 on page 1809 that aims to model the relationship between education and annual earnings across gender.\footnote{The paper does not provide reproducible code, and our data from which we derive results differ by a few observations.}


\subsection{Assumptions on the Data and Analysts’ Knowledge}
We make some assumptions to deal with limitations to existing DP methodology. 
We assume there are no survey weights and, for queries other than counts, there are no empty subsets for categorical variables.
To the best of our knowledge, there are no existing methods satisfying $(\epsilon,\delta)$-DP or $(\xi, \rho)$-zCDP that handle survey weights as generated by most federal statistical agencies, including SOI, because those systems extensively use the confidential data to generate the final survey weights (e.g., non-response adjustments). For the second assumption, if someone using a validation server applies their analyses to an empty subset, the query will result in an error message. If released, this error message will inform the user that there are no observations in the interested subset, violating DP. We acknowledge this as a practical issue but addressing this problem is beyond the scope of our paper.


In order to calculate the sensitivity, we also assume the users input bounds for the continuous variables and provides a list of the levels in categorical variables. We assume users would derive these inputs using prior knowledge or publicly available synthetic data. In our experiments, we use the observed minimum and maximum as bounds for continuous variables and observed levels for categorical variables. Using this strategy to set the bounds and levels implies that in practice results may be worse than our findings show. Our approach also implies that if a method has poor performance, we do not expect that altering the bounds or levels would lead to improvements. However, the proposed procedure can serve as a starting point and as a way to choose methods. Future work can focus on identifying strategies for determining input parameters.

\section{Utility Metrics for Experimentation}\label{sec:utility}
We use several statistical measures to evaluate the methods for releasing statistics under DP. Because we use real sample data rather than simulated, we define utility as the empirical distance between the confidential estimates and the estimates from DP methods. We broadly categorize the metrics that: (1) measure the error of the DP statistic relative to the confidential estimate and (2) measure the similarity of inferences that are drawn using both the point estimates and uncertainty between the confidential and DP output.

The first category consists of computing the $l_1$ error of the simulated DP statistics with respect to the confidential statistics. For the histograms and quantiles, where we produce multiple statistics, we compute the average $l_1$ error across multiple bins or quantiles.
The second category of utility metrics concerns statistics with uncertainty estimates, which are the mean and regression queries. The first metric we use to compare similarity of inference on the confidential data and the DP estimates is the confidence interval ratio (CIR). This is the ratio of width of the DP CI to the confidential CI, and this metric informs us how much increased uncertainty comes from using the DP estimate.

Along with CIR, we calculate the percentage of times across the simulated DP regression estimates that the following three criteria are all achieved: (1) the signs match between the confidential data and DP estimates, (2) the significance (assuming 0.05 level) match between the confidential and DP estimates, and (3) the CIs for the confidential and DP estimates overlap. We term this metric as sign, significance, and overlap (SSO) match. The SSO match percentage across simulated DP estimates gives a heuristic metric for how frequently the DP estimates are expected to provide similar inference to the confidential estimates. As we will show, there are cases where DP estimates give a high SSO match percentage, but have dramatically wider CIs than the non-DP estimates. We provide CIR and SSO together to describe the empirical utility that aligns with data users' analyses. If the CIR is close to 1 and the SSO match percentage is high, then the DP estimates from these mechanisms enable inferences similar to those based on the confidential estimates.

These measures are similar conceptually to the confidence interval overlap (CIO) measure from \citet{karr2006framework}. However, we found the measure hard to interpret in the regression results for our study. Specifically, when the confidential and DP CIs do not overlap, the CIO value is negative. But, interpreting this negative value is unclear. Conversely, when the DP CI completely encompasses the confidential CI, the CIO cannot be less than 0.5, regardless of how wide the DP CI becomes. The results shown in the Supplementary Material for the estimate of simple means and CIs give the CIO values, but we do not include them for the regression results.

Although the CIR and SSO measures are less familiar to a statistical audience, they give an interpretive way of assessing the inferential differences between the confidential and DP outputs. In a practical setting, this evaluation tells us how the DP methods could impact the public policy decisions for a particular tax case study.

\section{Results of the Case Studies}\label{sec:summary-results}
Due to space, we only provide the regression queries results here. We show results of the evaluations of the DP methods for tabular, mean, and quantile estimates in the Supplementary Materials (along with results on the SOI PUF data). When testing all our methods, we run all DP methods 100 times for each of $\epsilon =\{0.01, 0.1, 0.5, 1, 5, 10, 15, 20\}$,  $\delta=\{10^{-3},10^{-7},10^{-10}\}$, and use a bootstrap size of 10 and 25 for the BHM for means. We direct readers interested in reproducing our results to our GitHub repo,\footnote{GitHub repo website, https://github.com/UrbanInstitute/formal-privacy-comp-appendix}
which contains the code, public data, and complete results from our study.

\subsection{Regression Results}\label{sec:regresion-results}
We evaluate the methods listed in Table \ref{tab:reg} for producing DP estimates of linear regression coefficients. 
In order to improve readability, we present results only for selected methods and values of $\epsilon$ and $\delta$, depending on the range of the results. Certain methods clearly outperformed other methods, so we only present the results for the two best performing methods. 
Results  for additional levels of the privacy parameters, additional methods, and SOI PUF results can be found in the Supplementary Materials and our GitHub repo.


We estimate the relationship between education and annual earnings by gender\footnote{In the paper, separate models are fit for men and women. We present results here for the model fit on women.} with log annual earnings as the dependent variable. This is often called a Mincer model. The model contains six coefficients, including the intercept, and we display the results for two here. The ``Non-White" coefficient is for a binary indicator variable, and on the confidential data it has a point estimate close to zero with a 95\% confidence interval that covers zero. The ``Years of Education" coefficient is for a count variable, and it has a strong positive effect with a tighter 95\% confidence interval that does not cover zero. We select these two coefficients from the six to highlight the differences in utility depending on the strength of the relationship between the predictor and the outcome in the confidential data. The other three non-intercept predictors included in the model are quite skewed, so they present a different challenge. As can be seen in the Supplementary Material, their utility is very poor due to their large sensitivity (based on their sample bounds). We emphasize that the results presented here should be seen as a ``good case."

\begin{figure}[!htb]
    \centering
    \includegraphics[width=6in]{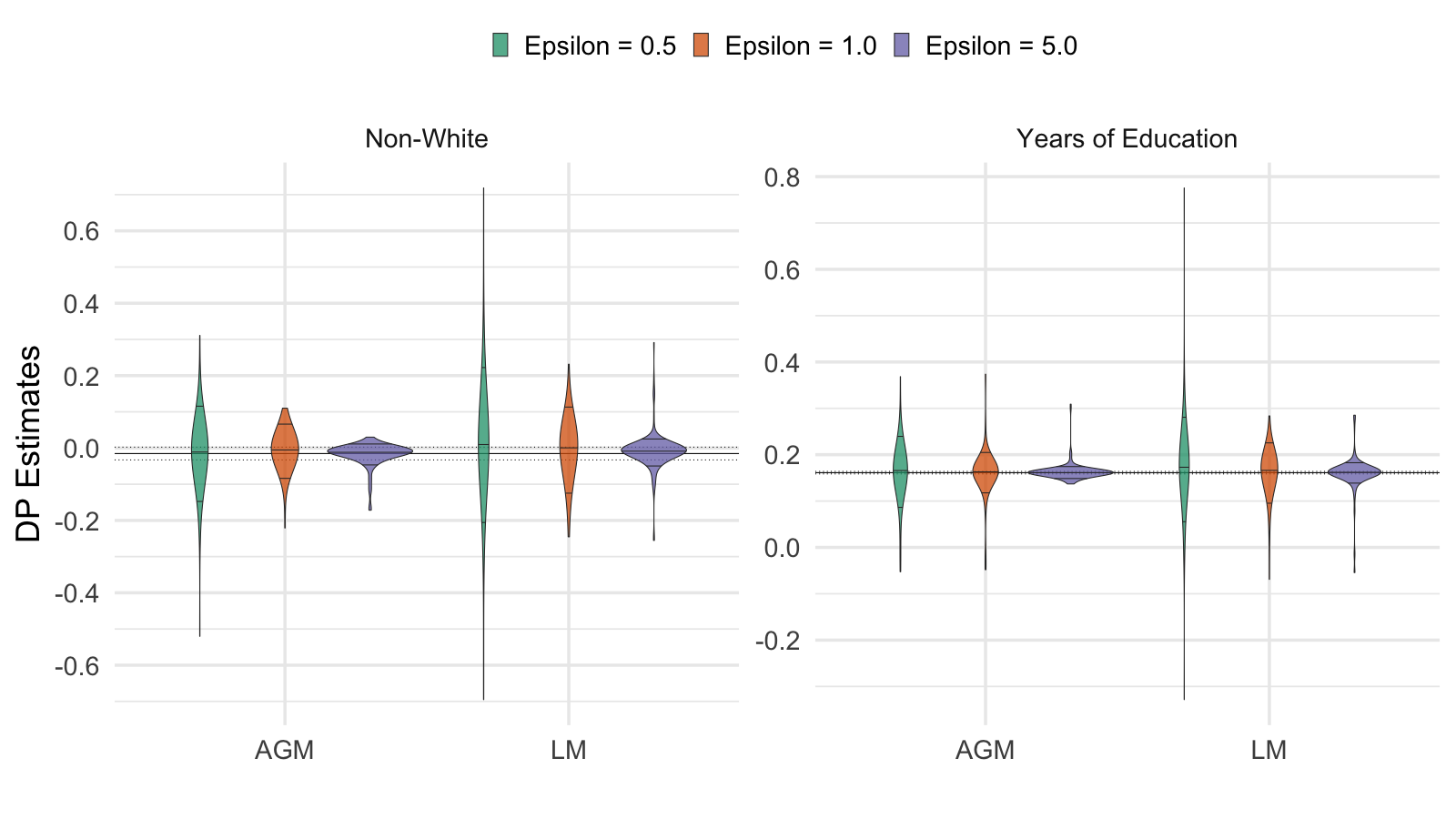}
    \caption{Distribution of simulated DP estimates for regression coefficients. Confidential estimates (black horizontal lines) and 95\% confidence intervals (dotted lines) are shown. $\delta = 10^{-7}$ for AGM.}
    \label{fig:reg_abs_bias_all}
\end{figure}

Figure \ref{fig:reg_abs_bias_all} shows the distributions of simulated DP outputs from the Analytic Gaussian Mechanism (\textit{AGM}) and Laplace Mechanism (\textit{LM}) for three levels of $\epsilon$. We see that distributions of the outputs for both DP mechanisms get tighter around the confidential estimates (solid black lines) as $\epsilon$ grows. But, a large proportion of the simulated estimates fall outside the confidential 95\% confidence intervals (dotted lines) even for $\epsilon = 5$. This result suggests that the noise added to protect privacy from these mechanisms is larger than what we would expect from the sampling variability. We remind readers that these results are from the two best performing methods. While many simulated queries returned for this model contain satisfactory results, there is a reasonably high probability that the query could return estimates that are far from the confidential estimates.

\begin{table}[!htp]
    \def\arraystretch{1}
    \caption{Summary of the regression results. Confidential (non-DP) estimate values for each coefficient are shown. 5th, 95th, quantiles, and $l_1$ Error for select coefficients, methods, and values of $\epsilon$. $\delta = 10^{-7}$ for AGM.}
    \label{tab:cps_reg_results_error}
    \centering
    \scriptsize
    \small
    \begin{tabular}{C{4.75cm} | C{1.5cm} | C{1.6cm} | C{1.75cm} | C{1.75cm} | C{1.5cm}}
        \hline
            \textbf{Coefficient (Estimate)} & \textbf{Epsilon} & \textbf{Method} &  \textbf{DP 5\%} & \textbf{DP 95\%} & \textbf{$l_1$} \\
        \hline
            \multirow{ 6}{*}{Non-White (-0.015)} & 0.5 & \textit{AGM} & -0.144 & 0.105 &  0.064 \\
            & 0.5 & \textit{LM} & -0.233 & 0.225 & 0.109 \\\cline{2-6}
            & 1 & \textit{AGM} & -0.070 & 0.065 & 0.037 \\
            & 1 & \textit{LM} & -0.128 & 0.108 & 0.059 \\\cline{2-6}
            & 5 & \textit{AGM} & -0.084 & 0.009 & 0.018 \\
            & 5 & \textit{LM} & -0.063 & 0.021 & 0.025 \\
            \hline
            \hline
            \multirow{ 6}{*}{Years of Education (0.161)} & 0.5 & \textit{AGM} & 0.083 & 0.240 & 0.038 \\
            & 0.5 & \textit{LM} & 0.018 & 0.282 & 0.060 \\\cline{2-6}
            & 1 & \textit{AGM} & 0.113 & 0.211 & 0.026 \\
            & 1 & \textit{LM} & 0.077 & 0.212 & 0.032 \\\cline{2-6}
            & 5 & \textit{AGM} & 0.149 & 0.174 & 0.008 \\
            & 5 & \textit{LM} & 0.124 & 0.181 & 0.016 \\
        \hline
    \end{tabular}
\end{table}


Table \ref{tab:cps_reg_results_error} summarizes the empirical error from these estimates with respect to the non-DP estimate. The table shows the 5th and 95th quantiles of the simulated outputs and the $l_1$ error for three levels of $\epsilon$. We see that the AGM outperforms LM and both mechanisms provide smaller errors as $\epsilon$ increases. The size of the errors are smaller for the ``Years of Education'' coefficient both in absolute terms and relative to the confidential estimate. Interpreting the coefficients as odds ratios may help better understand the potential impact on a user's inference if they were provided DP estimates. Based on the confidential estimates, an increase for each additional one year of education corresponds to an estimated relative increase in the annual earnings by 17.5\%. Under the AGM, the 90\% simulated range of this estimated relative increase in annual earnings for each additional one year of education is 12.0\% to 23.5\% for $\epsilon = 1$ and 16.1\% to 19.0\% for $\epsilon = 5$.

Next, we evaluate the estimated private CIs compared to the confidential estimates. We show the AGM and LM results for the two different estimation methods for producing CIs, asymptotic and bootstrap. We use the CIR to compare how much the width of the CI increases due to the DP mechanism, and we measure the percentage of simulated DP estimates that match SSO. 
If both the CIR is close to 1 and SSO match percentage is high, then the DP estimates from these mechanisms enable inferences similar to those based on the confidential estimates.

\begin{table}[!htp]
    \def\arraystretch{1}
    \caption{Summary of the private CI estimates. Median and 90th percentile (in parenthesis) CIR and percentage of SSO match for select coefficients, methods, and values of $\epsilon$. Both methods for estimating CI shown. $\delta = 10^{-7}$ for AGM.}
    \label{tab:cps_reg_results_ci}
    \centering
    \scriptsize
    \small
    \begin{tabular}{ C{3.5cm} | C{1.5cm} | C{1.6cm}| C{2cm} | C{1.25cm} | C{2cm} | C{1.25cm}}
            \hline
            \multicolumn{3}{c|}{} & \multicolumn{2}{|c|}{\textbf{Asymptotic}} & \multicolumn{2}{|c}{\textbf{Bootstrap}} \\
            \hline
            \textbf{Coefficient} & \textbf{Epsilon} & \textbf{Method} & \textbf{CIR} & \textbf{SSO\%} & \textbf{CIR} & \textbf{SSO\%} \\
            \hline
            \multirow{ 6}{*}{Non-White} & 0.5 & \textit{AGM} & 1.04 (1.15) & 8\% & 9.74 (47.4) & 53\% \\
            & 0.5 & \textit{LM} & 1.11 (1.26) & 8\% & 13.0 (72.3) & 42\% \\\cline{2-7}
            & 1 & \textit{AGM} & 1.02 (1.08) & 19\% & 4.50 (19.8) & 54\% \\
            & 1 & \textit{LM} & 1.03 (1.16) & 7\% & 7.20 (25.5) & 46\% \\\cline{2-7}
            & 5 & \textit{AGM} & 1.00 (1.02) & 53\%  & 1.78 (17.5) & 82\% \\
            & 5 & \textit{LM} & 1.01 (1.03) & 43\%  & 2.03 (28.9) & 71\% \\
            \hline
            \hline
            \multirow{ 6}{*}{Years of Education} & 0.5 & \textit{AGM} & 1.03 (1.15) & 4\% & 32.1 (102) & 75\% \\
            & 0.5 & \textit{LM} & 1.08 (1.21) & 9\% & 42.7 (180) & 58\% \\\cline{2-7}
            & 1 & \textit{AGM} & 1.02 (1.07) & 24\% & 14.8 (65.8) & 86\% \\
            & 1 & \textit{LM} & 1.04 (1.13) & 15\% & 24.7 (101) & 79\% \\\cline{2-7}
            & 5 & \textit{AGM} & 1.00 (1.02) & 62\% & 4.43 (26.6) & 93\% \\
            & 5 & \textit{LM} & 1.01 (1.03) & 42\% & 6.03 (81.8) & 84\% \\
            \hline
    \end{tabular}
\end{table}

Table \ref{tab:cps_reg_results_ci} shows the results using the asymptotic and bootstrap methods for estimating private CIs. Using the asymptotic estimator, we see that the CIR is always close to 1, which we would expect because this method does not take the noise from the privacy mechanism into account. For SSO, when $\epsilon$ is 0.5 or 1, we see that the match rate is below 10 or 25\%, respectively. At $\epsilon = 5$, the rate is roughly 50-60\% for the AGM. Conversely, the bootstrap CIs are much wider, but they provide a higher SSO match percentage. When $\epsilon = 5$, the AGM using bootstrap CIs gives 80 to 90\% for the SSO match, and the private CI is roughly 2 to 4 times wider than the confidential CI.

These results demonstrate that using either AGM or LM can produce regression estimates with either similar CI width or reasonable SSO percentage match but not both. At $\epsilon = 5$, the results may be good enough for some users. However, compared to other DP applications, data users would need to spend a high value of privacy loss for a single model. We note again that this is a ``good case." The other coefficients in the model and the SOI PUF results show worse utility (provided in the Supplementary Material). 

Both asymptotic and bootstrap interval estimates can be provided without incurring additional privacy loss, so the validation server might give both to the user. This scenario would provide both an estimate of the variation due to the privacy mechanism and the variation in the confidential estimate without noise. Further work is needed to understand how data users might utilize both these sources of information together.

To help visualize the simulated point and CIs estimates under the two different methods, we plot all 100 simulated estimates with 95\% CIs for the AGM. Each case is color-coded for whether they match sign, significance, and overlap (blue), or if they do not match at least one (red). We remind the readers that for the ``Non-White" coefficient, the confidential estimate (black) has a 95\% CI that covers zero, whereas ``Years of Education" coefficient does not. This distinction is relevant, because, in the former case, SSO match is achieved with private CIs that cover zero. In the latter case, SSO is achieved with private CIs that do not cover zero. We also plot only the 90 best simulated estimates in order to aid plot readability.

\begin{figure}[!htb]
    \centering
    \includegraphics[width=6.5in]{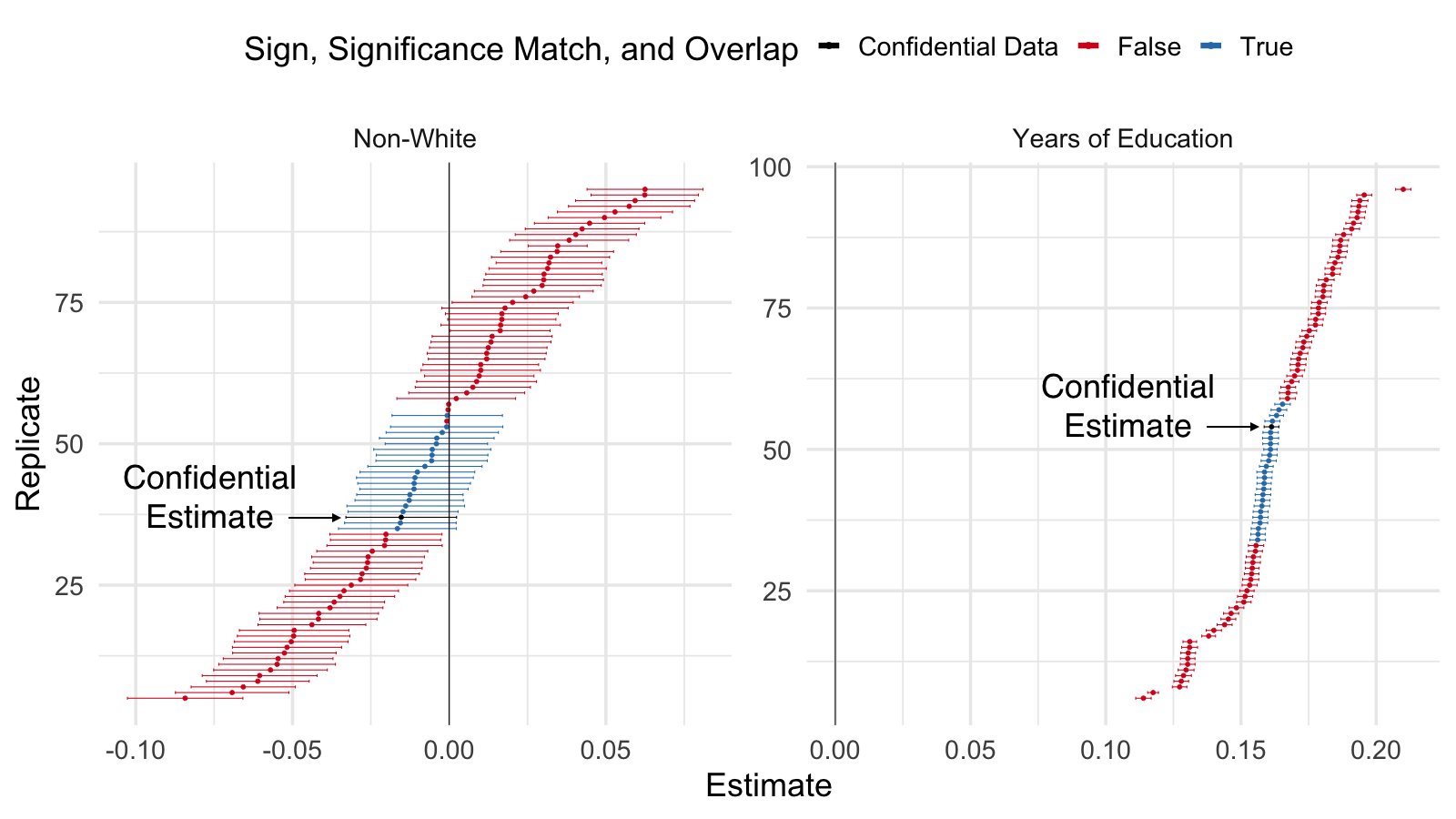}
    \caption{Regression results showing the 90 best (out of 100) simulated DP estimates and confidence intervals for the Analytic Gaussian Mechanism and $\epsilon = 1$. Results shown for two coefficients with confidence intervals estimated using the \textbf{asymptotic} approach. $\delta = 10^{-7}$ for AGM.}
    \label{fig:reg_catepillar_asymptotic}
\end{figure}

Figure \ref{fig:reg_catepillar_asymptotic} shows the simulated estimates with the asymptotic CIs. In this case, we see, as expected, that the intervals are all roughly equal length to the confidential interval. As the point estimates move away from the confidential estimate the intervals stop overlapping. One might argue that this is not a large problem for many of the ``Years of Education" DP estimates. However, the ``Non-White" DP estimates that are significantly positive or negative would lead to substantially different inference from the confidential estimates.

\begin{figure}[!htb]
    \centering
    \includegraphics[width=6.5in]{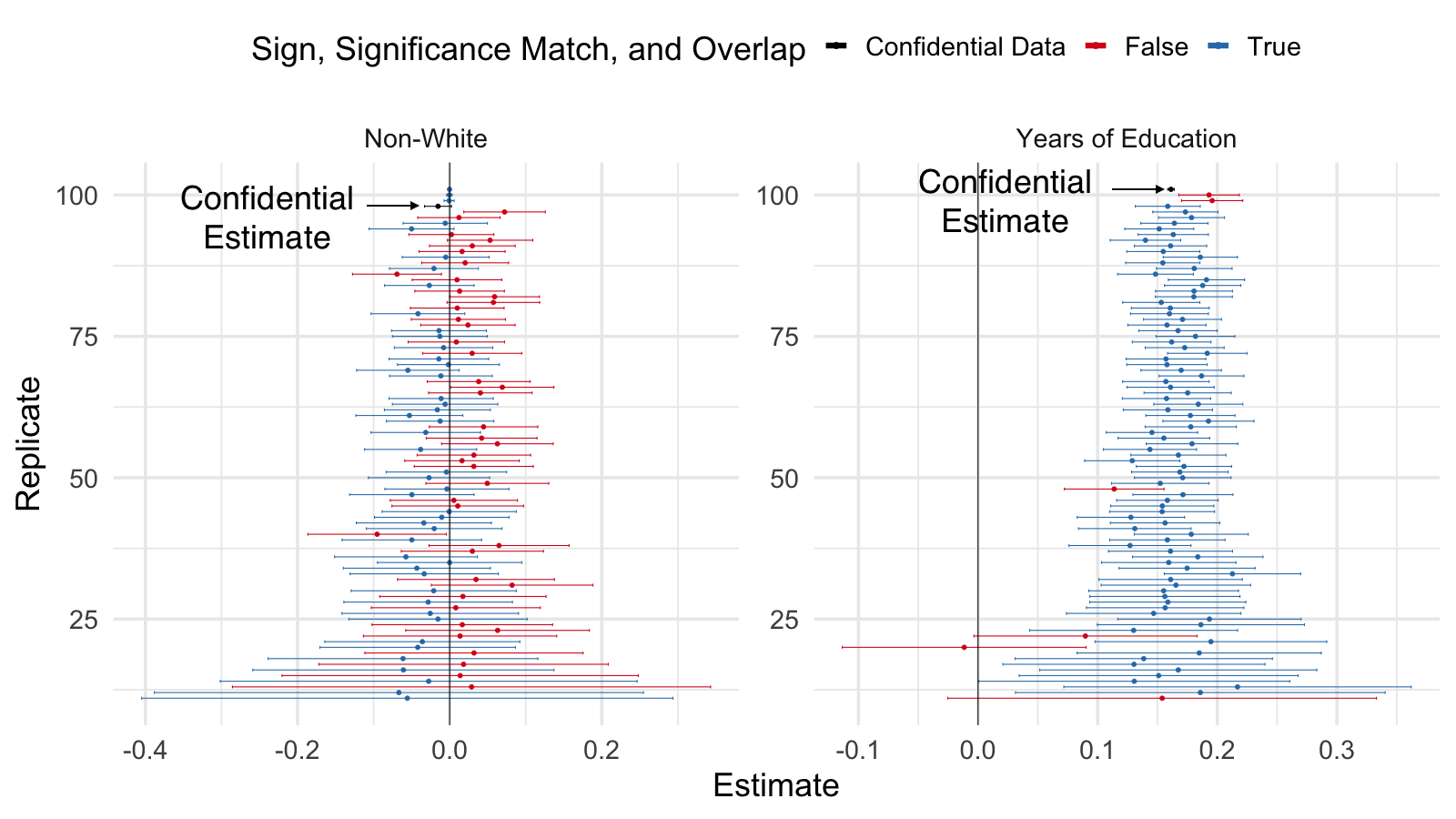}
    \caption{Regression results showing the 90 best (out of 100) simulated DP estimates and confidence intervals for the Analytic Gaussian Mechanism and $\epsilon = 1$. Results shown for two coefficients with confidence intervals estimated using the \textbf{bootstrap} approach. $\delta = 10^{-7}$ for AGM.}
    \label{fig:reg_catepillar_boot}
\end{figure}

Figure \ref{fig:reg_catepillar_boot} shows the DP estimates using the bootstrap CIs. In contrast to the previous figure, the noisy CIs are much wider than the confidential estimate. They almost all overlap or cover the confidential CI, since they take into account the extra variation induced by the DP mechanism. For the ``Non-White" coefficient many estimates switch sign, but the intervals typically cover zero. In the case of ``Years of Education", we see that, even with the increased uncertainty, most of the estimates are still significant, because the confidential estimate is strongly significant. As was seen in Table \ref{tab:cps_reg_results_ci}, the intervals are much wider, but the DP estimates using bootstrap CIs are more likely to match SSO.

\section{Conclusions from the Case Study}\label{sec:conclusion}
In this paper, we survey and test the feasibility of the latest DP methods for summary statistics and regression analyses on real tax data (see the Supplemental materials for the summary statistics results). To the best of our knowledge, this is the first comprehensive evaluation of these DP methods for practical applications within a validation server framework for real-world data. 
We find that regression methods require at least $\epsilon = 5$ for full inference. Practical applications would likely require either larger sample sizes or allocating more $\epsilon$ on every query in order to return estimates with satisfactory levels of uncertainty. Based on our results, we identify a few challenges and avenues for future work.

\subsection{Challenges}\label{subsec:challenges}
We find that existing DP regression methods are limited in applicability and often add high levels of noise. The only methods we found that met our inclusion criteria perturb the sufficient statistics prior to computing the noisy regression estimates. We identify three primary shortcomings with perturbing the sufficient statistics. First, the privacy budget must be split $\frac{(k + 1)\cdot(k + 2)}{2}$ times for $k$ coefficients. Second, when perturbing the sufficient statistics, the added noise becomes multiplicative rather than additive. Third, most methods are designed for normal data, so, when applied to skewed data, the results worsen significantly because of increased sensitivity.

Besides methodological issues, we encounter challenges with coding the various algorithms. For instance, obtaining code for applications and ensuring it is error-free presents challenges. Although we do not expect academics to provide production-ready code, code often fell short of standards for reproducibility like those outlined by the American Economic Association.\footnote{American Economic Association's Data and Code Availability Policy. Accessed: 2023-05-18. https://www.aeaweb.org/journals/data/data-code-policy} Research code was often messy, hard to read, and difficult to alter for our use cases. In some cases, the issues are minor and were easily addressed through consultation with the authors. In other cases, we could not use the method due to a lack of functioning code. When no code was available and the author(s) did not respond, we attempted to code the methods ourselves. But, in some cases, the manuscript does not provide enough information for us to implement the method and is thus excluded from the feasibility study. These issues emphasize the importance of open-source code to facilitate wider use and acceptance of DP algorithms in practice. For example, OpenDP from Harvard University is developing a suite of open-source software tools to implement DP methods.
These platforms are still under development, but they may offer improved solutions in the future.

\subsection{Future Work}\label{subsec:future}
One vital area for improvement is developing DP algorithms for data that are not Gaussian. Many of the methods we test performed well in their original papers, because the authors tested on well-behaved or normally distributed data. Real data are often skewed, such as the 2012 SOI PUF and CPS ASEC data, resulting in these same methods performing poorly.

Another area for improvement is developing DP mechanisms for regression with a focus on estimating the uncertainty of the estimates. Many data privacy experts create DP regression methods to output accurate predictions. But, as seen in Section \ref{sec:regresion-results}, these methods perform poorly either by returning very large confidence intervals or by not reporting the standard error at all. This appears to be a significant gap in the DP literature that must be addressed.

In addition to improving DP methods for regression-based statistical inference, future work should focus on expanding research to address other important economic use cases. Given the challenges with directly querying regression coefficient and interval estimates based on perturbed sufficient statistics, an entirely different approach may be needed. Determining these alternative approaches is beyond the scope of this paper, but our findings suggest that DP methods for regression have significant barriers to overcome before they can enable accurate statistical inference.
We hope this study provides the statistical data privacy community with a better understanding of current DP methods for summary statistics and regression analyses, including the capabilities, limitations, and challenges.

\section*{Acknowledgments}
\noindent \textbf{Funding:} This research was funded by the Alfred P. Sloan Foundation [G-2020-14024] and National Science Foundation National Center for Science and Engineering Statistics [49100420C002].

\noindent \textbf{Collaborators:} We would like to thank our collaborators at SOI, especially Barry Johnson and Victoria Bryant, for their amazing support. We also thank our stellar validation server project team, consisting of Leonard Burman, John Czajka, Surachai Khitatrakun, Graham MacDonald, Rob McClelland, Livia Mucciolo, Madeline Pickens, Silke Taylor, Kyle Ueyama, Doug Wissoker, and Noah Zwiefel. Thank you to Gabriel Morrison for reviewing our code. Finally, we thank our advisory board for their advice and support. The members are John Abowd, Jim Cilke, Jason DeBacker, Nada Eissa, Rick Evans, Dan Feenberg, Max Ghenis, Nick Hart, Matt Jensen, Barry Johnson, Ithai Lurie, Ashwin Machanavajjhala, Shelly Martinez, Robert Moffitt, Amy O'Hara, Jerry Reiter, Emmanuel Saez, Wade Shen, Aleksandra Slavkovi\'c, Salil Vadhan, and Lars Vilhuber.

\noindent \textbf{Author Contributions:} \textbf{AFB:} Conceptualization, Formal analysis, Methodology, Software, Writing - original draft, Writing - review \& editing; \textbf{ARW:} Conceptualization, Data Curation, Formal analysis, Resources, Software, Validation, Writing - original draft, Writing - review \& editing; \textbf{JS:} Conceptualization, Formal analysis, Methodology, Software, Visualization, Writing - original draft, Writing - review \& editing; \textbf{CMB:} Conceptualization, Funding acquisition, Project administration, Resources, Software, Validation, Writing - original draft, Writing - review \& editing

\newpage
\begin{center}
\large
\textbf{Supplemental Materials for\\ ``A Feasibility Study of Differentially Private Summary Statistics and Regression Analyses with Evaluations on Administrative and Survey Data"}\\
\vspace{12pt}

\end{center}

This file contains the Supplemental Materials to accompany the paper ``A Feasibility Study of Differentially Private Summary Statistics and Regression Analyses with Evaluations on Administrative and Survey Data."

\section{Setting the Privacy Loss Budget}
The total privacy budget and the best way to allocate portions of the privacy budget to each query (or user) are areas for discussion that we do not attempt to answer in this paper. The scientific community still has no general consensus on what value of $\epsilon$ should be used for practical implementation. Early DP research focused on $\epsilon$ values that were less than or equal to one and suggested that an epsilon of two or three would release too much information \citep{dwork2008survey}. Researchers have also proposed other technical interpretations relating to hypothesis testing \citep{wasserman2010statistical} or odds-ratios \citep{machanavajjhala2008privacy} to interpret and set limits on $\epsilon$. 

More recently, many privacy researchers working in practical applications frame the decision as a social choice question. They interpret the parameter as a way to quantify the trade-off between accuracy and worst-case privacy loss \citep{abowd2019economic}. This means that privacy experts should explain to policymakers ways of interpreting the privacy and utility trade-off. Ultimately stakeholders will need to set the privacy parameter in ways that are relevant to their contexts, such as in keeping with laws that apply to the data. 

Accordingly, many practical applications of DP use large values of $\epsilon$ (by theoretical standards). In 2008, for example, privacy researchers applied $(\epsilon,\delta)$-DP method with values at $(8.6,10^{-5})$ to release a synthetic version of the OnTheMap data, a United States commuter dataset \citep{machanavajjhala2008privacy}. More recently, in 2020, Google's COVID-19 Mobility Reports used $2.64$-DP for the daily reports; a total of $79.22$-DP monthly \citep{aktay2020google}. In the same year, LinkedIn revealed their LinkedIn's Audience Engagement API that protected LinkedIn members' content engagement data, which used $(\epsilon,\delta)$-DP with daily values of $(0.15,10^{-10})$ or $(34.9,7\times 10^{-9})$ for monthly queries \citep{rogers2020linkedin}. In 2021, the Census Bureau used privacy loss of $\rho$ = 2.56 on person-level redistricting data and $\rho$ = 0.07 on unit-level redistricting data for a total privacy loss budget on this product of $\rho$ = 2.63. This converts to $\epsilon=17.41$ and $\delta= 10^{-10}$.\footnote{The U.S. Census Bureau set the final privacy loss budget on June 9, 2021, which can be found at https://www.census.gov/newsroom/press-releases/2021/2020-census-key-parameters.html. The specific privacy budget allocation can be found at https://www2.census.gov/programs-surveys/decennial/2020/program-management/data-product-planning/2010-demonstration-data-products/ppmf20210608/2021-06-08-privacy-budgetallocation.pdf.}

Given these examples and the evolving understanding of $\epsilon$, we explore a wide range of $\epsilon$ values in this paper based on values seen in theoretical work and practical applications. By doing so, we gain a better sense of the methods' feasibility at different levels of $\epsilon$. Specifically, we examine the effects of $\epsilon$ and $\delta$ on accuracy within the context of our study, which we hope will contribute to the conversation about how to set the privacy loss budget.

When considering the total privacy loss budget, it is important to understand that choosing the value of $\epsilon$ for any single query is sensitive to other factors, such as the sample size, the total number of desired queries, and the size of the population from which data are sampled. Those questions concern the overall framework of a fully implemented validation server, and we do not seek to answer those questions in this paper. We make this point to note that our goal is simply to identify which algorithms are desirable, if any, based on their relative privacy and utility trade-offs, rather than their absolute trade-offs.

\section{Extended Review of Differentially Private Regression Methods}\label{subsec:reg-review}
As stated in the main text, DP methods for regression can be classified according to the outputs they produce: (1) point estimates only, (2) point estimates and interval estimates, and (3) other outputs related to regression analysis, such as diagnostic plots. In this section, we cover categories (1) and (3).

Most DP algorithms fall into the first category, returning only noisy point estimates. Methods in this category frequently rely on objective-perturbation-based approaches, such as the functional mechanism \citep{chaudhuri2011differentially, zhang2012functional, fang2019regression, gong2019differential}. These approaches provide DP coefficients estimates by maximizing a perturbed version of the objective function. We can obtain the perturbed versions of the objective function by either adding noise to the function or using a polynomial representation of the function, where we added noise to the polynomial coefficients.

Other DP point estimate methods borrow ideas from robust statistics. For example, \cite{avella2020privacy} proposes $(\epsilon, \delta)$-differential private methods to obtain robust estimators for various problems, including regression analysis. \cite{chen2020median} developed $\epsilon$-DP methods for median regression, which can be seen as a robust version of ordinary least squares based regression. Finally, some DP point estimate approaches add noise to sufficient statistics \citep{wang2018revisiting}. For a list of DP  methods for simple linear regression, we refer the reader to  \cite{alabi2020differentially}. 

Next, we review recent contributions that release outputs distinct from point and interval estimates. Even though these methods are beyond the current scope of the validation server, they provide key information for regression analysis and could be fruitful additions to future versions of the validation server.

\cite{barrientos2019differentially} proposed an $\epsilon$-DP method to perform hypothesis testing for single coefficients. This approach has the advantage of not requiring the response and predictors to be bounded. However, the privacy loss budget may be costly when performing hypothesis testing for multiple coefficients using this method. The Bayesian procedure by \cite{amitai2018differentially} has similar capabilities and limitations, since it targets specific summaries of the posterior distributions of the regression coefficients, such as tail probabilities.

Residual analysis is another important task in regression and a crucial tool for model validation when finding solutions to problems, such as heterogeneity of errors and lack of linearity. \cite{chen2016differentially} describe $\epsilon$-DP methods that release a private version of the residuals versus fitted values plot. Model selection is also key for regression analysis, and \cite{lei2016differentially} has developed an $(\epsilon, \delta)$-DP algorithm for this purpose. Other contributions related to regression analysis involve algorithms developed for regularized regression, such as Lasso \citep{talwar2015nearly,dandekar2018differential}.

\subsection{Details on Differentially Private Bootstrap Regression}\label{subsec:boot}
\cite{ferrando2020general} proposed a DP bootstrap-based algorithm that adds noise to the sufficient statistic for linear model $Y = X\beta+\e$, where $Y$ is the vector representing the observations for the response, $X$ is the design matrix, and $\e$ is the vector of independent and identically distributed normal errors. We assume that the reported noisy statistic is of the form
$$
S_H^r = S + H - rI_{(p+1)\times (p+1)} 
$$
where $S=[X,Y]^t[X,Y]$ is the sufficient statistic for linear model $Y = X\beta+\e$, the matrix $H$ denotes the noise added to achieve differential privacy using either the Laplace, Analytic Gaussian, or Wishart mechanism, and the $r$ is defined as discussed in Section 3.4.2. For the Normal and Spherical Laplace mechanisms, we define $S_H^r$ as the already-regularized resulting matrix obtained from \cite{wang2019differentially}'s Algorithm 2. We denote $P_H$ as the corresponding probability distribution of $H$ under a given mechanism. To define $H$ for the Normal and Spherical Laplace mechanisms, we use a symmetric version of the added noise as in \cite{wang2019differentially}'s Algorithm 2, i.e., we define $H = (\tilde H + \tilde H^t)/2$ where the entries of $\tilde H$ are drawn from the corresponding normal or spherical Laplace distribution. The up-to-date version of the algorithm proposed by \cite{ferrando2020general} assumes that $r$ is equal to zero and only considers the Laplace mechanism. This algorithm also assumes that the sample size $n$ is known. This is something we cannot assume and, for this reason, we replace the sample size by a DP version of it. Notice that when the intercept is included in the model and represented by the first column of $X$, a privatized version of the sample size is available at the entry (1,1) of $S_H^r$. If the intercept is not included, users will need to spend part of their privacy budget querying this quantity. The algorithm below summarizes the employed algorithm after modifications and adaptations.

\begin{algorithm} \caption{DP bootstrap-based regression}
\begin{algorithmic}[1]
\Input \hfill  \break 
\begin{tabular}{ll}
     $S_H^r$: regularized noisy sufficient statistic  &  $P_H$: probability distribution of $H$\\
     $B$: number of bootstrap samples & $p$: number of regression coefficients\\
     &
\end{tabular}
\State $\widehat{n} \gets S_H^r\texttt{[1,1]}$ \Comment{assuming the intercept is part of the model}
\State $\widehat{X^TX} \gets S_H^r\texttt{[1:p,1:p]}$ \Comment{submatrix of $S_H^r$ with \texttt{1:p}$=(1,\ldots,p)$.}
\State $\widehat{X^TY} \gets S_H^r\texttt{[1:p,p+1]}$
\State $\widehat{Y^TY} \gets S_H^r\texttt{[p+1,p+1]}$
\State $\hat \beta \gets (\widehat{X^TX})^{-1}\widehat{X^TY} $
\State $\hat \sigma^2 \gets (\widehat{Y^TY}-(\widehat{X^TY})^T(\widehat{X^TX})^{-1}\widehat{X^TY})/(\widehat{n}-p-1)$
\For{$b$ in $\{1,\ldots,B\}$}
\State Sample $h$ from $P_H$
\State Sample $\widetilde{X^T\boldsymbol{u}}$ from $N(0,\hat n \hat \sigma^2 \widehat{X^TX} ) $
\State $\tilde \beta_b  \gets (\widehat{X^TX})^{-1}(\widehat{X^TX} - h\texttt{[1:p,1:p]})\hat \beta +
(\widehat{X^TX})^{-1}(\widetilde{X^T\boldsymbol{u}}+h\texttt{[1:p,p+1]})$
\EndFor
\Output $\tilde \beta_1,\ldots,\tilde \beta_B$
\end{algorithmic}
\end{algorithm} 

\subsection{Sensitivity Calculations}\label{subsubsec:reg_meth}

To compute the sensitivity of $S$ for the Laplace mechanism described in Section 3.4.2, we assume that the response and predictors are bounded---a common assumption in differential privacy. Without loss of generality, we assume that the response and predictors take values $[0,1]$. Because $S$ is a symmetric matrix and under the previous assumption, its sensitivity is upper-bounded by the number of entries in and above the diagonal, i.e., $(p+1)(p+2)/2$, where $p$ is the number of regression coefficients. While using the upper bound $(p+1)(p+2)/2$ as sensitivity is a valid strategy, it is particularly inefficient in the presence of categorical predictors. To improve this upper bound, we implement some strategies when categorical predictors are part of the analysis. We assume that all categorical predictors are included in the model as dummy variables and fixing one level as the reference. The implemented strategies are listed below: 

\begin{enumerate}
    \item One or more entries in the diagonal of $S$ are identical to entries in the off-diagonal. Hence, we only need to add noise to the unique entries.
    
    \item If the categorical predictor has more than two levels, then some entries of $S$ will be exactly zero, eliminating the need to add noise to those entries.
    
    \item If the categorical predictor has more than two levels and the model has an intercept, then we multiply the set of the corresponding dummy variables by the column of ones in $X$ (representing the intercept) results in a vector that counts the number of ones in the dummy variables. The vector is equal to a contingency table that counts the number of observations at each level, leaving out the reference one. Although this vector has a dimension greater than one, its sensitivity is equal to one. Hence, we take advantage of this fact to reduce the sensitivity of $S$.
    
    \item If the categorical predictor has more than two levels and the model has a numeric predictor, then we multiply the set of corresponding dummy variables by the column in $X$ representing the numerical predictor results in a vector of partial sums. Each entry of this vector is equal to the summation of numerical predictor values across the observations that share the same level as the categorical predictor. Similar to the previous item, while this vector has a dimension greater than one, its sensitivity is equal to one. Hence, we take advantage of this fact to reduce the sensitivity of $S$.
\end{enumerate}

We follow the same strategy to define the sensitivity for the Analytic Gaussian mechanism and the approach proposed by \cite{brawner2018bootstrap}. For the Wishart mechanism, the sensitivity is equal to the upper bound for $l_2$-norm of the rows in $[X,Y]$. We therefore only consider item 2, i.e., an entry of $S_H$ is set to be equal to zero if the same entry in $S$ is zero by definition. 

We also realize that the magnitude of the sensitivity depends on the upper and lower bounds of the variables (response and predictors) involved in the analysis. Since the variables are often in different scales, users can easily face scenarios where the magnitude of the sensitivity is highly dominated by, for example, a single variable. This single variable could have an interval length (upper minus lower bound) that is much larger than those of the remaining variables. If users ignore this issue under such scenario, the mechanism will add too much noise to those summaries involving the remaining variables. To avoid this problem, we first use the provided bounds to rescale all variables to lie in the $[0,1]$ interval. We then implement the DP method and compute the point and interval estimates of the regression coefficients. Finally, we use the provided bounds again to scale back and report the estimates in the original scale.


\section{CPS Experimental Results}\label{sec:supp_cps}
In this section, we present tables and figures for the DP tabular, mean, and quantile results. We also provide additional results for the regression coefficients not shown in the main paper.

\subsection{Histograms for Tabular Summaries}\label{subsec:hist}
\begin{table}[htbp]
    \def\arraystretch{1.1}
    \caption{Summary of the DP tabular methods reviewed for the study}
    \label{tab:tab_supplemental}
    \centering
    \small
    \begin{tabular}{ C{3cm} | C{2.75cm} | C{4cm} | C{5cm} }
        \hline
            \multicolumn{4}{c}{\textit{\textbf{Tabular Statistics}}} \\
        \hline
            \textbf{Method} & \textbf{Privacy Definition} & \textbf{Off-the-Shelf vs. Hand-Coding} & \textbf{Selected for Case Study} \\
        \hline
            Laplace mechanism & $\epsilon$-DP & off-the-shelf via \verb;R; and \verb;Python; code on GitHub & Yes \\
        \hline
            Gaussian mechanism & $(\epsilon,\delta)$-DP & off-the-shelf via \verb;R; and \verb;Python; code on GitHub & Yes \\
        \hline
    \end{tabular}
\end{table}
We test the Laplace and Gaussian mechanisms from Table \ref{subsec:hist} for providing counts of individuals within income categories.\footnote{We considered making separate queries for each income bin, but the results were strictly worse than making one query for the whole histogram.} We calculate the maximum error over all cumulative sums as a percentage of the sum over the whole histogram. We also compute the mean error over all cumulative sums. These two metrics allow us to interpret the expected error given how a tax researcher might use the queried histograms. Discerning bunching or jumps in the distribution is useful for developing more complex models and relies on understanding the cumulative distribution function. The maximum relative error tells us the biggest change across the distribution in the total percentage of individuals who fall below or above a certain cutoff, informing the jumps. Therefore, the mean error tells us the average error across all subsets of the histogram and indicates where bunching occurs. 
\begin{table}[!htp]
    \def\arraystretch{1}
    \caption{Summary of the histogram results. Median simulated values for the errors are provided and 99th percentile simulated values are in parenthesis. Select values of $\epsilon$ provided. For Gaussian mechanism, $\delta = 10^{-7}$}.
    \label{tab:cps_hist_results_supplemental}
    \centering
    \scriptsize
    \small
    \begin{tabular}{ C{4cm} | C{2cm} | C{3.25cm} | C{3cm}}
        \hline
            \textbf{Method} & \textbf{Epsilon} & \textbf{Max Relative Error} & \textbf{Mean Error} \\
        \hline
            Laplace mechanism & 0.01 & 5.19\% (11.5\%) & 30.7\% (47.1\%) \\
            Gaussian mechanism & 0.01 & 13.3\% (28.5\%) & 84.1\% (123.2\%) \\
            \hline
            Laplace mechanism & 0.1 & 0.54\% (1.15\%) & 3.11\% (4.94\%) \\
            Gaussian mechanism & 0.1 & 2.11\% (4.69\%) & 13.2\% (19.0\%) \\
            \hline
            Laplace mechanism & 1 & 0.06\% (0.12\%) & 0.30\% (0.48\%) \\
            Gaussian mechanism & 1 & 0.21\% (0.42\%) & 1.32\% (1.84\%) \\
            \hline
            Laplace mechanism & 5 & 0.01\% (0.02\%) & 0.02\% (0.07\%) \\
            Gaussian mechanism & 5 & 0.04\% (0.09\%) & 0.25\% (0.38\%) \\
        \hline
    \end{tabular}
\end{table}

\begin{figure}[!htb]
    \centering
    \includegraphics[width=\textwidth]{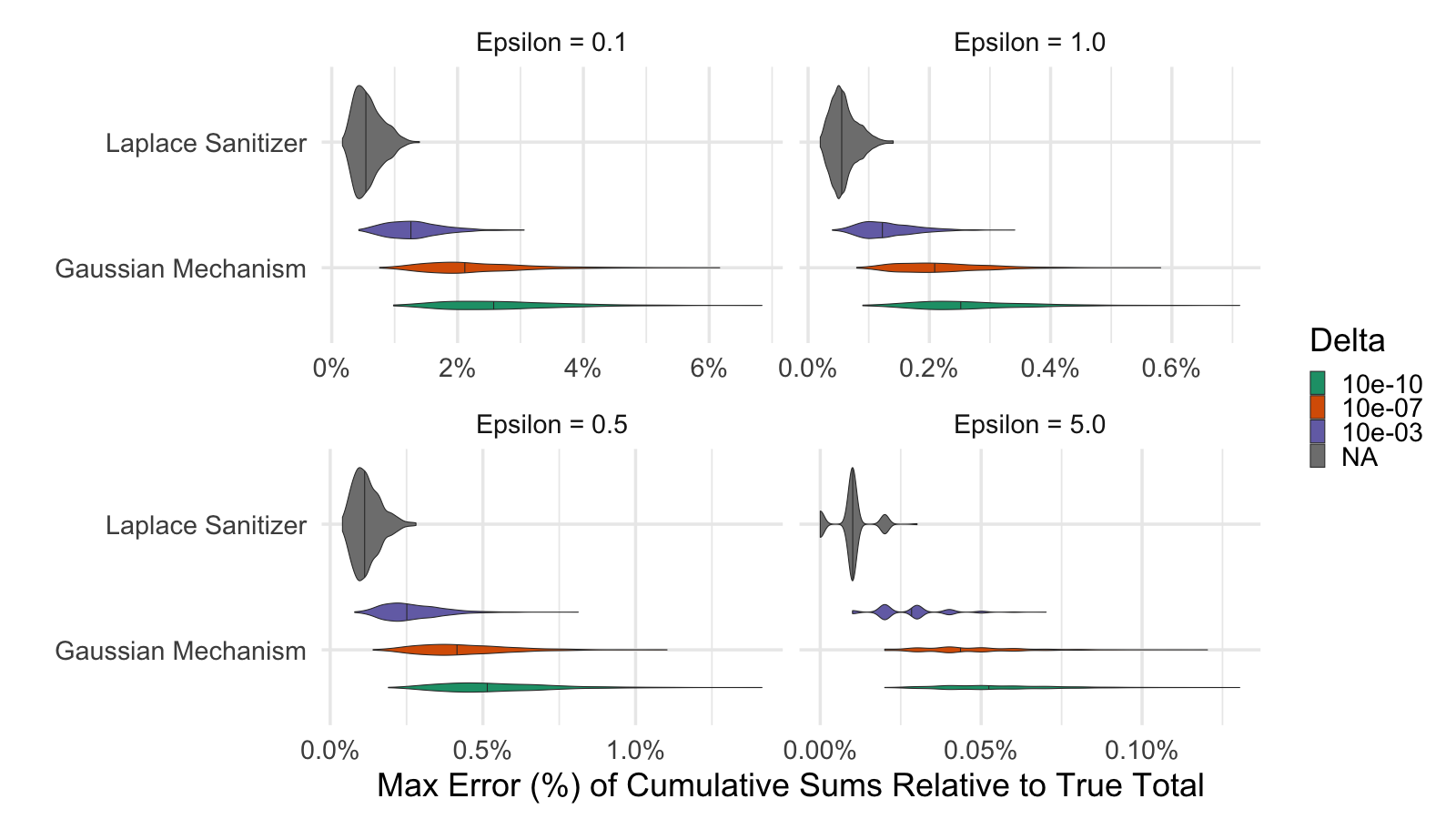}
    \caption{Income histogram results: max relative error for two methods and varying levels of $\epsilon$ and $\delta$.}
    \label{fig:inc_hist_max_supplemental}
\end{figure}

\begin{figure}[!htb]
    \centering
    \includegraphics[width=\textwidth]{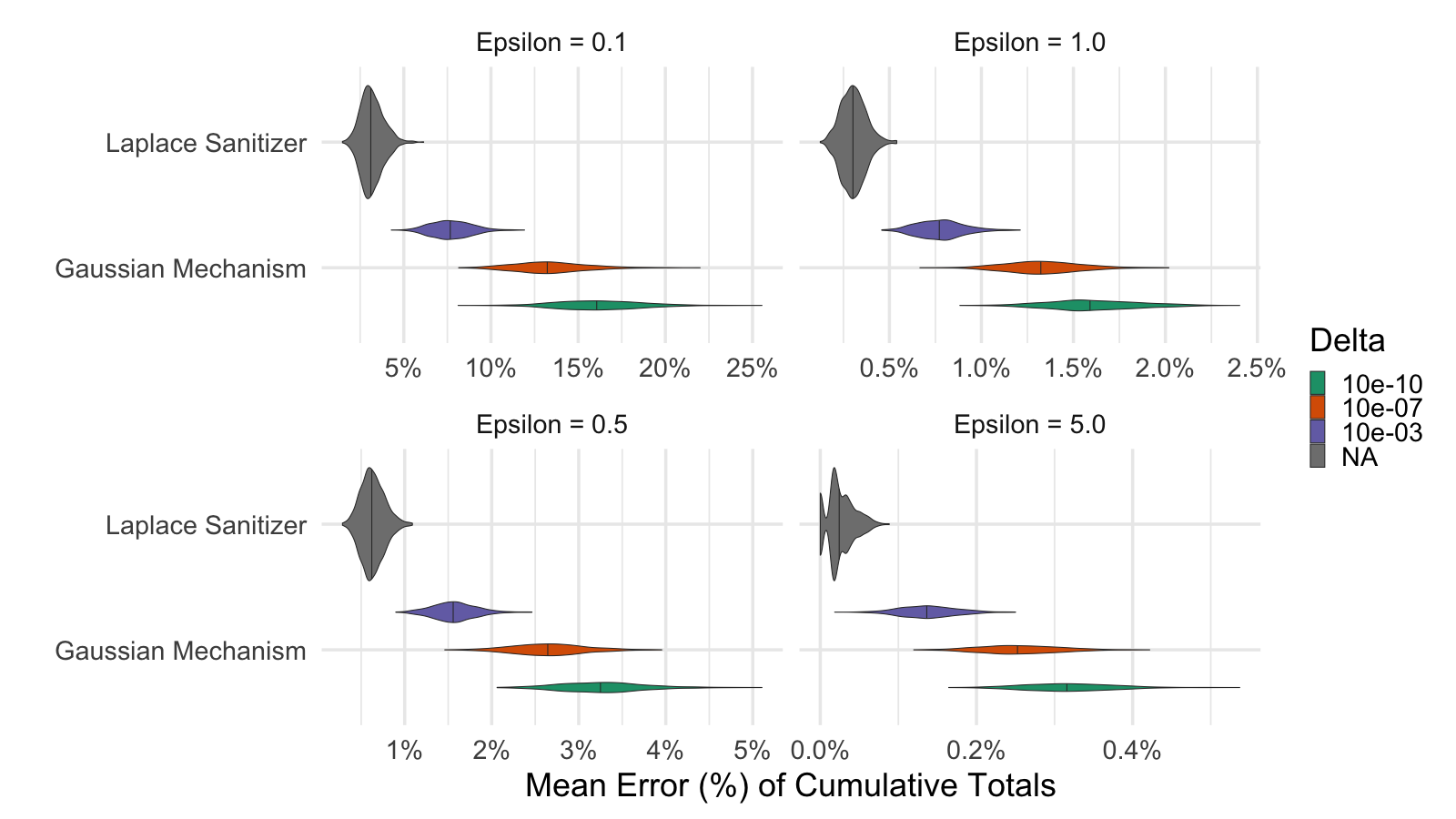}
    \caption{Income histogram results: mean cumulative sums error for two methods and varying levels of $\epsilon$ and $\delta$.}
    \label{fig:inc_hist_mean_supplemental}
\end{figure}

Table \ref{tab:cps_hist_results_supplemental} and Figures \ref{fig:inc_hist_max_supplemental} and \ref{fig:inc_hist_mean_supplemental} provide the results. Overall, the utility is high. We see that the max relative error and mean error are below 0.5\% for $\epsilon = 1$ in 99\% of the cases for the Laplace mechanism and below 5\% for $\epsilon = 0.1$. With lower error and a stronger privacy guarantee, the Laplace mechanism is preferable.

\subsection{Quantiles}\label{subsec:quant}
\begin{figure}[!htb]
    \centering
    \includegraphics[width=\textwidth]{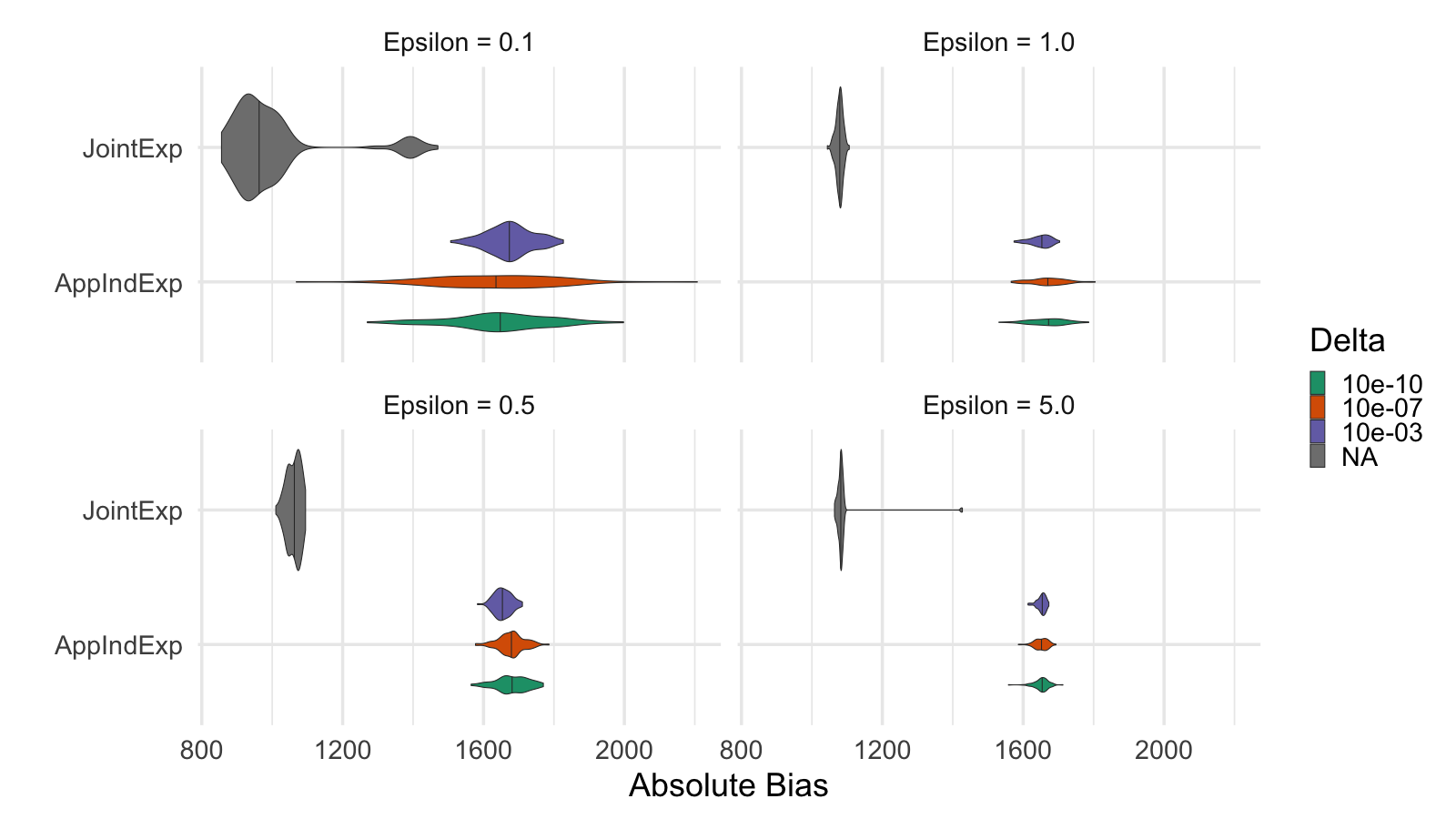}
    \caption{Average absolute bias across quantiles for two methods and varying levels of $\epsilon$ and $\delta$.}
    \label{fig:quantile_abs_supplemental}
\end{figure}

\begin{table}[htbp]
    \def\arraystretch{1.1}
    \caption{Summary of the DP quantile statistics reviewed for the study}
    \label{tab:quant_supplemental}
    \centering
    \small
    \begin{tabular}{ C{3cm} | C{2.75cm} | C{4cm} | C{5cm} }
        \hline
            \multicolumn{4}{c}{\textit{\textbf{Quantile Statistics}}} \\
        \hline
            \textbf{Method} & \textbf{Privacy Definition} & \textbf{Off-the-Shelf vs. Hand-Coding} & \textbf{Selected or Not Selected for Case Study} \\
        \hline
             \textit{AppIndExp} \citep{smith2011privacy,gillenwater2021differentially} & $(\epsilon,\delta)$-DP & off-the-shelf via \verb;Python; code on GitHub & Yes \\
        \hline
            \textit{JointExp} \citep{gillenwater2021differentially} & $\epsilon$-DP & off-the-shelf via \verb;Python; code on GitHub & Yes \\
        \hline
            \textit{Smooth} \citep{nissim2007smooth} & $(\epsilon,\delta)$-DP & off-the-shelf via \verb;Python; code on GitHub & Yes \\
        \hline
            \textit{Concentrated Smooth} \citep{gillenwater2021differentially} & CDP & off-the-shelf via \verb;Python; code on GitHub & No, requires fine tuning which is not realistic for our application \\
        \hline
            Propose-Test-Release \citep{dwork2009differential} & $(\epsilon,\delta)$-DP & No code publicly available & No, requires fine tuning which is not realistic for our application \\
        \hline
            \cite{tzamos2020optimal} & $\epsilon$-DP & No code publicly available & No, requires fine tuning which is not realistic for our application  \\
        \hline
    \end{tabular}
\end{table}
We test \textit{AppIndExp}, \textit{JointExp}, and \textit{Smooth} from Table \ref{tab:quant_supplemental} for estimating quantiles. Using the CPS income variable, we estimate the income deciles from 10\% to 90\%. Researchers could use these values for producing tax policy tables\footnote{\url{https://www.taxpolicycenter.org/statistics/household-income-quintiles}}. We compute the mean error across quantiles and the mean error across the ordered differences between quantiles. The latter would be of interest to researchers if they are trying to understand income disparities. Table \ref{tab:cps_quant_results_supplemental} and Figure \ref{fig:quantile_abs_supplemental} shows the results. 

\begin{table}[!htp]
    \def\arraystretch{1}
    \caption{Summary of the quantile results. Median simulated values for the mean errors across 9 quantiles are provided and 99th percentile simulated mean errors are in parenthesis. Select values of $\epsilon$ provided. For \textit{AppIndExp} and \textit{Smooth}, $\delta = 10^{-7}$}.
    \label{tab:cps_quant_results_supplemental}
    \centering
    \scriptsize
    \small
    \begin{tabular}{ C{3cm} | C{2cm} | C{3.25cm} | C{5.5cm}}
        \hline
            \textbf{Method} & \textbf{Epsilon} & \textbf{Mean Absolute Bias (\$)} & \textbf{Mean Ordered Differences Absolute Bias (\$)} \\
        \hline
            \textit{AppIndExp} & 0.01 & 3,136 (120,932) & 739 (118,697) \\
            \textit{JointExp} & 0.01 & 919 (2,518) & 88 (339) \\
            \textit{Smooth} & 0.01 & 553,500 (1,042,832) & 124,667 (124,667) \\
            \hline
            \textit{AppIndExp} & 0.1 & 1,644 (2,047) & 108 (216) \\
            \textit{JointExp} & 0.1 & 956 (1,421) & 115 (204) \\
            \textit{Smooth} & 0.1 & 119,401 (303,380) & 81,259 (124,667) \\
            \hline
            \textit{AppIndExp} & 1 & 1,670 (1,795) & 188 (221) \\
            \textit{JointExp} & 1 & 1,080 (1,100) & 179 (183) \\
            \textit{Smooth} & 1 & 2,589 (11,168) & 892 (5,532) \\
            \hline
            \textit{AppIndExp} & 5 & 1,652 (1,686) & 205 (221) \\
            \textit{JointExp} & 5 & 1,083 (1,423) & 210 (220) \\
            \textit{Smooth} & 5 & 1,652 (2,027) & 224 (294) \\
            \hline
        \hline
    \end{tabular}
\end{table}

We find that both \textit{AppIndExp} and \textit{JointExp} offer high utility, and they are generally preferable to \textit{Smooth} for all but the highest values of $\epsilon$. For $\epsilon < 5$, \textit{Smooth} performs much worse than the other two methods. This result is likely due to more extensive splitting of $\epsilon$ and $\delta$ required by the algorithm. We find \textit{JointExp} performs better than \textit{AppIndExp} at estimating the quantiles at lower levels of $\epsilon$. Both seem to quickly converge to their best solution by $\epsilon = 0.1$, but the error does not converge to zero even for higher levels of $\epsilon$. This suggests that the methods are not empirically unbiased and may be an artifact of the method for sampling from the Exponential mechanism, which is implemented in the source code.

For a practical validation server implementation, \textit{AppIndExp} and \textit{JointExp} both appear sufficient to return accurate quantile estimates. The choice likely depends on if the system deploys pure DP or approximate DP. We do not recommend that a system use \textit{Smooth} unless queries are made with very high levels of $\epsilon$. Additionally, we find that \textit{AppIndExp} returns more equally biased results for each quantile because they are drawn independently. On the other hand, \textit{JointExp} can return biased results, which are more biased for some quantiles than others. In our application, the quantiles follow an exponential trend. \textit{JointExp} returns more accurate results for the lowest and highest quantiles but returns less accurate results for those in-between. It may be preferable to choose one or the other algorithm, depending on the application.

\subsection{Estimates of the Mean with Confidence Intervals}\label{subsec:mean}

\begin{table}[htbp]
    \def\arraystretch{1.1}
    \caption{Summary of the DP mean confidence interval methods reviewed for the study}
    \label{tab:mean_ci_supplemental}
    \centering
    \small
    \begin{tabular}{ C{3cm} | C{2.75cm} | C{4cm} | C{5cm} }
        \hline
            \multicolumn{4}{c}{\textit{\textbf{Confidence Intervals for the Mean}}} \\
        \hline
            \textbf{Method} & \textbf{Privacy Definition} & \textbf{Off-the-Shelf vs. Hand-Coding} & \textbf{Selected or Not Selected for Case Study} \\
        \hline
             \cite{brawner2018bootstrap} & zCDP $\leftrightarrow$ $(\epsilon, \delta)$-DP & off-the-shelf via \verb;R; code on GitHub & Yes \\
        \hline
            \cite{d2015differential} & $\epsilon$-DP & off-the-shelf via \verb;R; code on GitHub & No, requires a priori bounds set on standard deviation \\
        \hline
            \cite{karwa2017finite} & $(\epsilon, \delta)$-DP & off-the-shelf via \verb;R; code on GitHub & No, requires a priori bounds set on standard deviation \\
        \hline
         \textit{COINPRESS} \citep{biswas2020coinpress} & zCDP & off-the-shelf via \verb;R; code on GitHub & No, requires a priori bounds set on variance/covariance \\
        \hline
        
            \textit{NOISYMAD} \citep{du2020differentially} & $\epsilon$-DP & off-the-shelf via \verb;R; code on GitHub & Yes \\
        \hline
            \textit{NOISYVAR} \citep{du2020differentially} & $\epsilon$-DP & off-the-shelf via \verb;R; code on GitHub & Yes \\
        \hline
            \textit{CENQ} \citep{du2020differentially} & $\epsilon$-DP & off-the-shelf via \verb;R; code on GitHub & Yes, for non-skewed data \\
        \hline
            \textit{MOD} \citep{du2020differentially} & $\epsilon$-DP & off-the-shelf via \verb;R; code on GitHub & Yes, for non-skewed data \\
        \hline
            \textit{SYMQ} \citep{du2020differentially} & $\epsilon$-DP & off-the-shelf via \verb;R; code on GitHub & Yes, for non-skewed data \\
        \hline
    \end{tabular}
\end{table}
We test \textit{NOISYMAD}, \textit{NOISYVAR}, and \textit{BHM} from Table \ref{tab:mean_ci_supplemental} for computing the mean income with an estimate of the 95\% confidence interval. Three additional methods listed in Table \ref{tab:mean_ci_supplemental} were not tested due to the heavy skew of the data. Our results indicate that they would only provide unbiased results for nearly perfectly Gaussian data.

\begin{table}[!htp]
    \def\arraystretch{1}
    \caption{Summary of the confidence interval measures for the mean statistic results. Median simulated values for the metrics are provided and 90th percentile (worse) simulated values are in parenthesis. Select values of $\epsilon$ provided. For BHM, $\delta = 10^{-7}$.}
    \label{tab:cps_mean_ci_results_supplemental}
    \centering
    \scriptsize
    \small
    \begin{tabular}{ C{4cm} | C{2cm} | C{3.25cm} | C{3cm}}
        \hline
            \textbf{Method} & \textbf{Epsilon} & \textbf{CIR} & \textbf{SSO} \\
        \hline
            \textit{NOISYMAD} & 0.1 & 1.30 (1.37) & 87\% (72\%) \\
            \textit{NOISYVAR} & 0.1 & 1.51 (1.63) & 82\% (70\%) \\
            \textit{BHM} & 0.1 & 7.14 (11.26) & 57\% (54\%) \\
            \hline
            \textit{NOISYMAD} & 0.5 & 0.73 (0.76) & 86\% (85\%) \\
            \textit{NOISYVAR} & 0.5 & 1.02 (1.07) & 97\% (91\%) \\
            \textit{BHM} & 0.5 & 1.92 (2.69) & 75\% (68\%) \\
            \hline
            \textit{NOISYMAD} & 1 & 0.70 (0.73) & 85\% (84\%) \\
            \textit{NOISYVAR} & 1 & 0.99 (1.04) & 98\% (94\%) \\
            \textit{BHM} & 1 & 1.28 (1.78) & 85\% (75\%) \\
            \hline
            \textit{NOISYMAD} & 5 & 0.69 (0.72) & 85\% (84\%) \\
            \textit{NOISYVAR} & 5 & 0.99 (1.04) & 99\% (98\%) \\
            \textit{BHM} & 5 & 0.99 (1.27) & 91\% (83\%) \\
        \hline
    \end{tabular}
\end{table}

\begin{figure}[!htb]
    \centering
    \includegraphics[width=\textwidth]{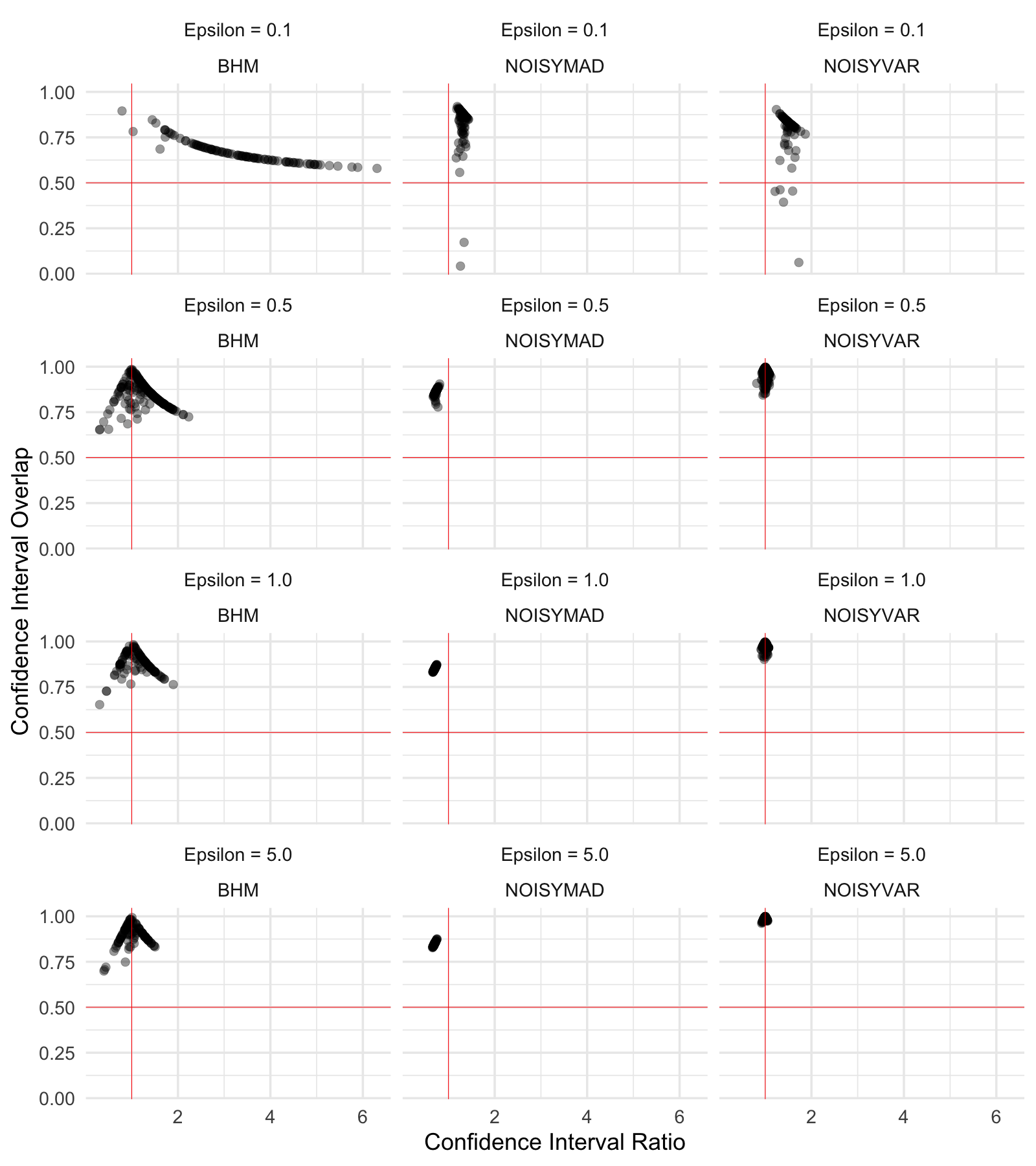}
    \caption{Mean income confidence intervals ratio and overlap results for three methods and varying levels of $\epsilon$. BHM Results Only Shown for $\delta = 0.01$.}
    \label{fig:mean_inc_ci_supplemental}
\end{figure}

Results are shown in Table \ref{tab:cps_mean_ci_results_supplemental} and Figure \ref{fig:mean_inc_ci_supplemental}. \textit{NOISYMAD} and \textit{NOISYVAR} perform similarly and both provide highly accurate statistics. We find that all three methods for means are approximately unbiased. \textit{BHM} performs comparably to the other two methods only when $\delta = 10^{-3}$ and when $\epsilon < 1$. Otherwise, the other two methods clearly outperform \textit{BHM}. For $\epsilon \geq 1$, the relative error is almost always less than $1\%$ for all methods.

We find more variation for the CI measures than point estimate bias, which reflects the differing approaches to estimating uncertainty. For $\epsilon \geq 0.5$, we find \textit{NOISYVAR} provides CIR values close to 1 and SSO values of 100\%. \textit{NOISYMAD} performs well for $\epsilon = 0.1$, but as $\epsilon$ becomes larger, the widths of the CIs produced by \textit{NOISYMAD} shrink and are consistently narrower than the confidential CIs. These results would produce overly confident inference for researchers performing hypothesis testing. For \textit{BHM}, the intervals are much wider especially for $\epsilon < 5$, resulting in high SSO match but also high CIR values that shrink towards 1 as $\epsilon$ grows more slowly that \textit{NOISYVAR}.

Figure \ref{fig:mean_inc_ci_supplemental} shows the simulated distributions of confidence interval overlap (CIO) and confidence interval ratio (CIR) values. We plot these two metrics against each other because taken together they greatly aid the interpretation of the results. The red lines plotted at $CIO = 0.5$ and $CIR = 1$ form four quadrants on the charts. We summarize the interpretation of each quadrant as follows: 
\begin{itemize}
    \item \textbf{Top left quadrant:} indicates noisy CIs that are more narrow than the confidential CI but mostly contained within the confidential CI.
    \item \textbf{Top right quadrant:} indicates noisy CIs that are wider than the confidential CI but mostly encompass the confidential CI.
    \item \textbf{Bottom left quadrant:} indicates noisy CIs that are both more narrow than the confidential CI and biased away from the confidential point estimate.
    \item \textbf{Bottom right quadrant:} indicates noisy CIs that are both wider than the confidential CIs and biased away from the confidential point estimate.
\end{itemize}

\subsection{Additional CPS Results for Regression Models}
In this section we provide results for the other estimated coefficients in the CPS model. Results for other methods we evaluate can be found in our GitHub repository.

The linear regression model we fit on the CPS data is specified as:
\begin{equation}
\begin{split}
    log(Annual\_Earnings) = \beta_0 + \beta_1 * I(NonWhite) + \beta_2 * Years\_of\_Education + \\
    \beta_3 * Potential\_Experience + \beta_4 * Potential\_Experience^2 + \beta_5 * Potential\_Experience^3
\end{split}
\end{equation}

\begin{figure}[!htb]
    \centering
    \includegraphics[width=6in]{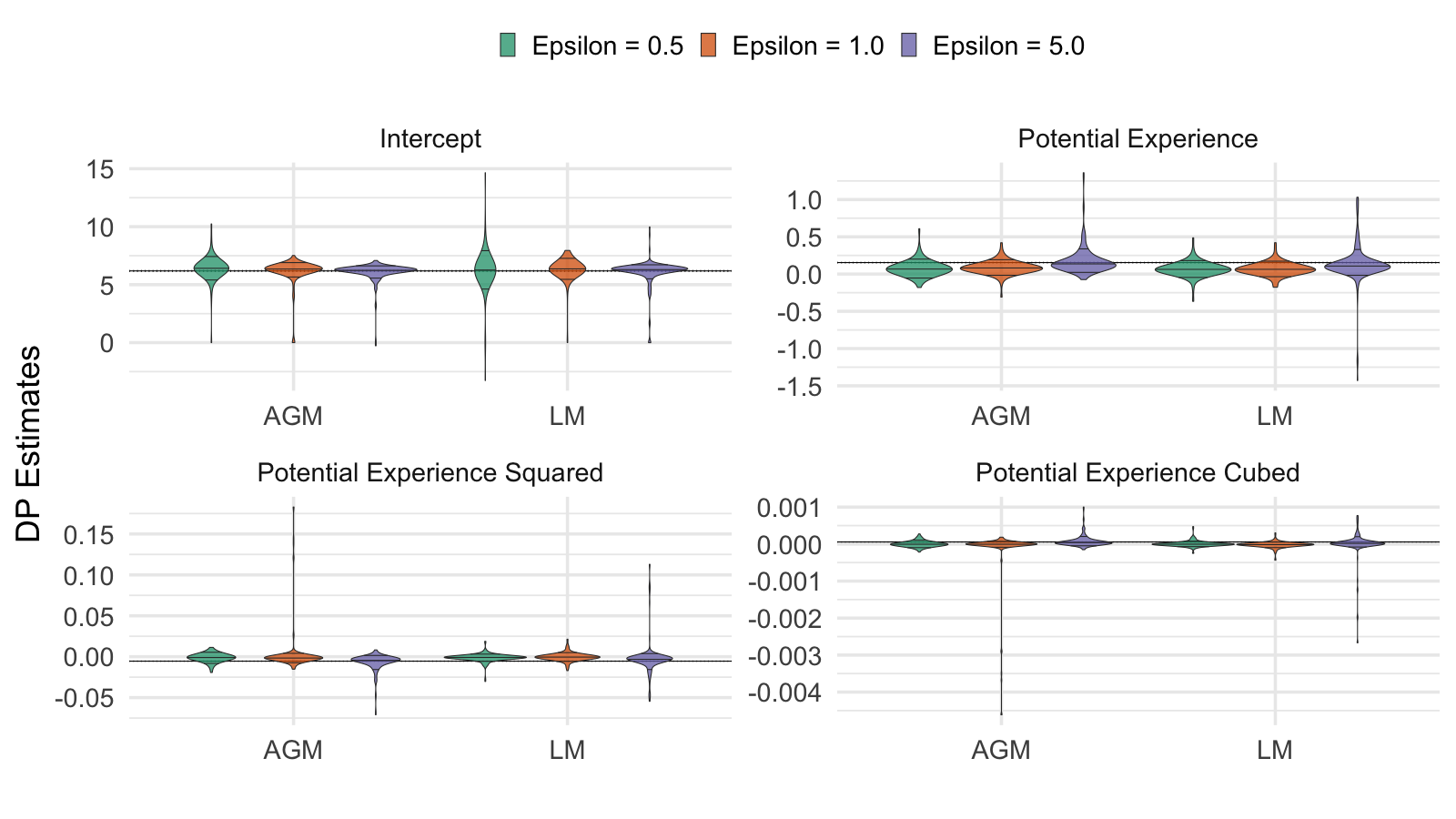}
    \caption{Distribution of simulated DP estimates for regression coefficients. Confidential estimates (black horizontal lines) and 95\% confidence intervals (dotted lines) are shown. $\delta = 10^{-7}$ for AGM.}
    \label{fig:reg_abs_bias_supplemental}
\end{figure}

\begin{figure}[!htb]
    \centering
    \includegraphics[width=6.5in]{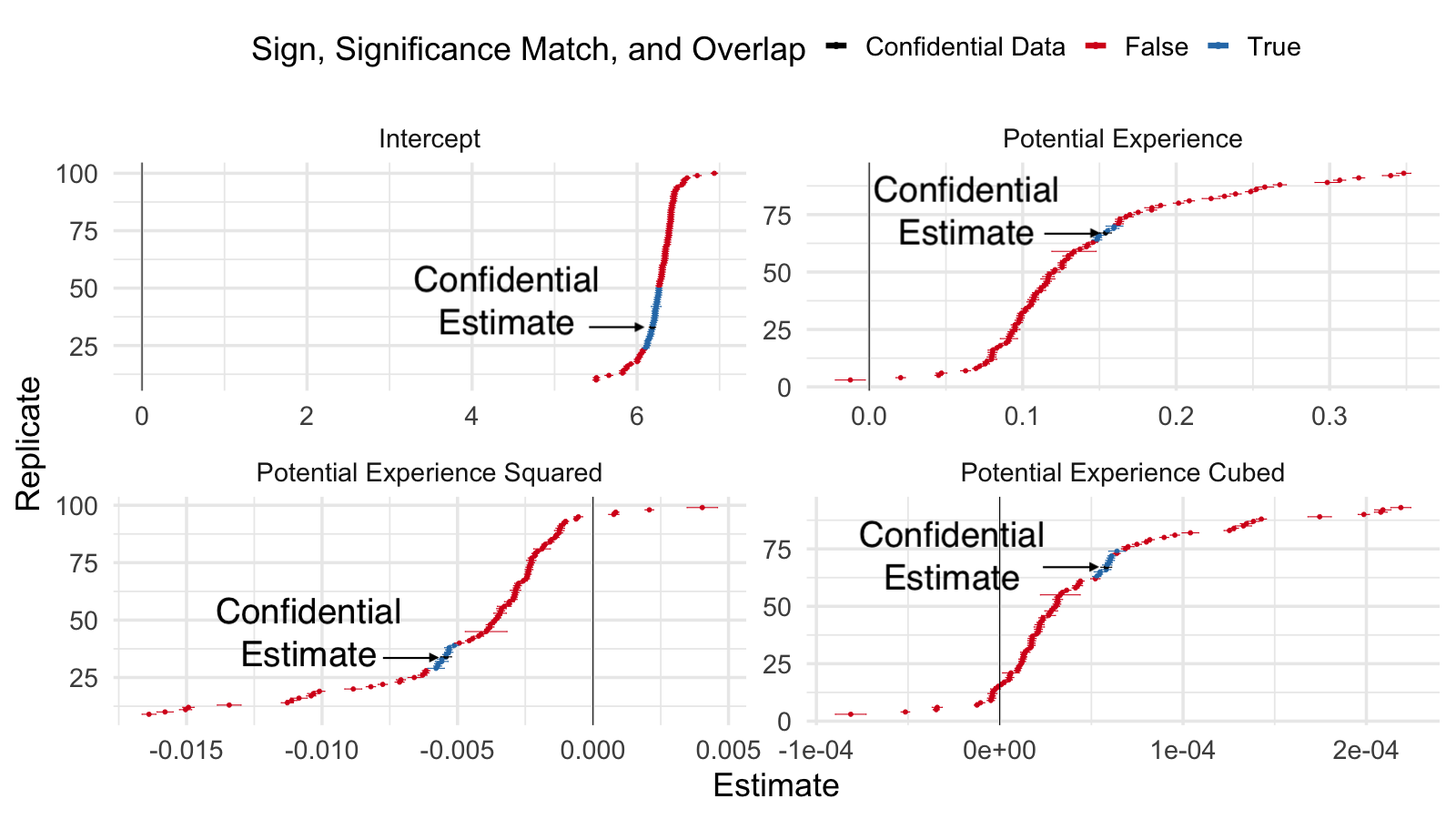}
    \caption{Regression results showing the 90 best (out of 100) simulated DP estimates and confidence intervals for the Analytic Gaussian Mechanism and $\epsilon = 5$. Results shown for four coefficients with confidence intervals estimated using the asymptotic approach. $\delta = 10^{-7}$ for AGM.}
    \label{fig:reg_catepillar_asymptotic_supplemental}
\end{figure}

\begin{figure}[!htb]
    \centering
    \includegraphics[width=6.5in]{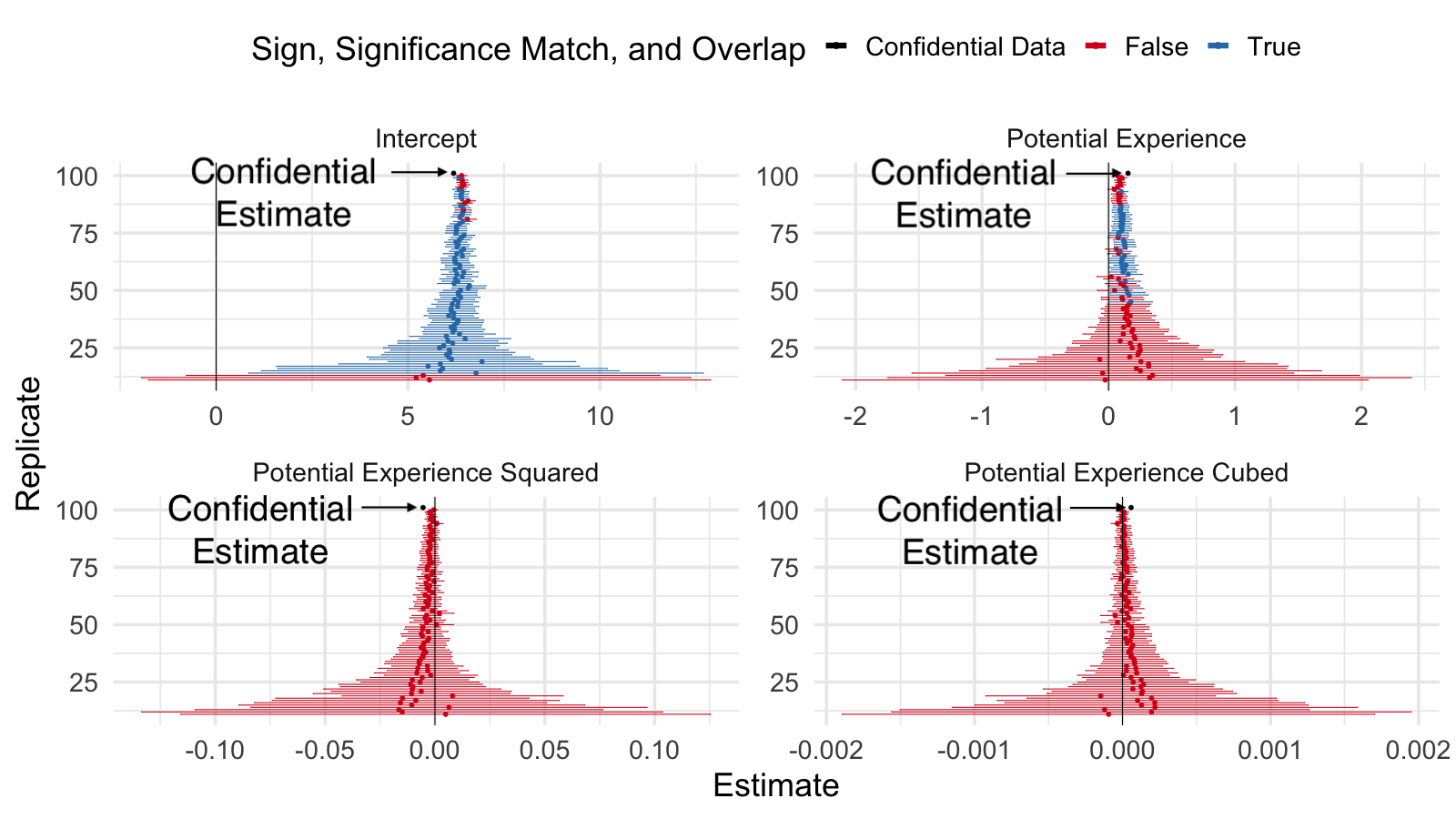}
    \caption{Regression results showing the 90 best (out of 100) simulated DP estimates and confidence intervals for the Analytic Gaussian Mechanism and $\epsilon = 5$. Results shown for four coefficients with confidence intervals estimated using the bootstrap approach. $\delta = 10^{-7}$ for AGM.}
    \label{fig:reg_catepillar_boot_supplemental}
\end{figure}

Figure \ref{fig:reg_abs_bias_supplemental} shows violin plots with the distribution of simulated estimates for the four coefficients not included in the main paper. Table \ref{tab:cps_reg_results_error_supplemental} provides the error rates and quantiles on the distribution of the DP estimates for the additional coefficients. We note that the results do not clearly improve as $\epsilon$ grows for the non-intercept coefficients. These variables are highly skewed, particularly when the square or cube is taken. Figures \ref{fig:reg_catepillar_asymptotic_supplemental} and \ref{fig:reg_catepillar_boot_supplemental} show the estimates and corresponding CIs estimated using the two different methods for the coefficients not included in the main paper. We see even at $\epsilon = 5$ that the results for the ``Potential Experience'' coefficients are still quite poor.

\begin{table}[!htp]
    \def\arraystretch{1}
    \caption{Summary of the regression results. Confidential (non-DP) estimate values for each coefficient are shown. 5th, 95th, quantiles, and L1 Error for select coefficients, methods, and values of $\epsilon$. $\delta = 10^{-7}$ for AGM.}
    \label{tab:cps_reg_results_error_supplemental}
    \centering
    \scriptsize
    \small
    \begin{tabular}{C{4.75cm} | C{1.5cm} | C{1.6cm} | C{1.75cm} | C{1.75cm} | C{1.5cm}}
        \hline
            \textbf{Coefficient (estimate)} & \textbf{Epsilon} & \textbf{Method} & \textbf{DP 5\%} & \textbf{DP 95\%} & \textbf{\textbf{$l_1$}} \\
        \hline
            \multirow{ 6}{*}{Intercept (6.19)} & 0.5 & \textit{AGM} & 5.35 &  7.44 & 0.596 \\
            & 0.5 & \textit{LM} & 4.80 & 8.00  & 0.943\\\cline{2-6}
            & 1 & \textit{AGM} & 4.72 & 6.83 & 0.513 \\
            & 1 & \textit{LM} & 5.60 & 7.53 & 0.488 \\\cline{2-6}
            & 5 & \textit{AGM} & 5.35 & 6.56 & 0.338 \\
            & 5 & \textit{LM} & 4.51 & 6.59 & 0.510 \\
            \hline
            \hline
            \multirow{ 6}{*}{Potential Exp. (0.154)} & 0.5 & \textit{AGM} & -4.65e-02 & 0.214 & 0.104 \\
            & 0.5 & \textit{LM} & -4.83e-02 & 0.199 & 0.104 \\\cline{2-6}
            & 1 & \textit{AGM} & -1.72e-02 & 0.221 & 8.66e-02 \\
            & 1 & \textit{LM} & -1.08e-02 & 0.175 & 0.101 \\\cline{2-6}
            & 5 & \textit{AGM} & 4.70e-02 & 0.472 & 9.38e-02 \\
            & 5 & \textit{LM} & 1.77e-02 & 0.0.518 & 0.129 \\
            \hline
            \hline
            \multirow{ 6}{*}{Potential Exp.$^2$ (-5.42e-03)} & 0.5 & \textit{AGM} & -1.01e-02 & 5.79e-03 & 5.46e-03 \\
            & 0.5 & \textit{LM} & -8.39e-03 & 3.63e-03 & 5.22e-03 \\\cline{2-6}
            & 1 & \textit{AGM} & -7.53e-03 & 5.00e-03 & 9.00e-03 \\
            & 1 & \textit{LM} & -5.54e-03 & 5.84e-03 & 5.76e-03 \\\cline{2-6}
            & 5 & \textit{AGM} & -2.27e-02 & 7.70e-04 & 5.28e-03 \\
            & 5 & \textit{LM} & -2.39e-02 & 2.50e-03 & 9.00e-03 \\
            \hline
            \hline
            \multirow{ 6}{*}{Potential Exp.$^3$ (5.81e-05)} & 0.5 & \textit{AGM} & -1.02e-04 & 1.48e-04 & 7.70e-05 \\
            & 0.5 & \textit{LM} & -6.66e-05 & 1.34e-04 & 7.24e-05 \\\cline{2-6}
            & 1 & \textit{AGM} & -9.21e-05 & 7.97e-05 & 1.76e-04 \\
            & 1 & \textit{LM} & -1.14e-04 & 6.28e-05 & 8.38e-05 \\\cline{2-6}
            & 5 & \textit{AGM} & -3.41e-05 & 3.08e-04 & 7.74e-05 \\
            & 5 & \textit{LM} & -6.01e-05 & 3.01e-04 & 1.48e-04 \\
            \hline
    \end{tabular}
\end{table}

\newpage
\section{SOI PUF Experimental Results}
In this section, we present tables and figures for the DP tabular, mean, quantile, and regression results on the private SOI PUF data. We use the same approach and utility metrics described for the CPS results in Section \ref{sec:supp_cps}.

\subsection{Histograms}
\begin{table}[!htp]
    \def\arraystretch{1}
    \caption{Summary of the SOI histogram results. Median simulated values for the errors are provided and 99th percentile simulated values are in parenthesis. Select values of $\epsilon$ provided. For Gaussian mechanism, $\delta = 10^{-7}$}.
    \label{tab:soi_hist_results_supplemental}
    \centering
    \scriptsize
    \small
    \begin{tabular}{ C{4cm} | C{2cm} | C{3.25cm} | C{3cm}}
        \hline
            \textbf{Method} & \textbf{Epsilon} & \textbf{Max Relative Error} & \textbf{Mean Error} \\
        \hline
            Laplace mechanism & 0.01 & 12.2\% (19.0\%) & 110\% (131\%) \\
            Gaussian mechanism & 0.01 & 16.6\% (22.2\%) & 127\% (147\%) \\
            \hline
            Laplace mechanism & 0.1 & 3.15\% (6.42\%) & 36.0\% (44.0\%) \\
            Gaussian mechanism & 0.1 & 7.61\% (13.8\%) & 87.4\% (100\%) \\
            \hline
            Laplace mechanism & 1 & 0.36\% (0.75\%) & 3.85\% (4.95\%) \\
            Gaussian mechanism & 1 & 1.34\% (2.63\%) & 16.8\% (19.8\%) \\
            \hline
            Laplace mechanism & 5 & 0.08\% (0.15\%) & 0.33\% (0.61\%) \\
            Gaussian mechanism & 5 & 0.28\% (0.57\%) & 3.28\% (3.92\%) \\
        \hline
    \end{tabular}
\end{table}

\begin{figure}[!htb]
    \centering
    \includegraphics[width=\textwidth]{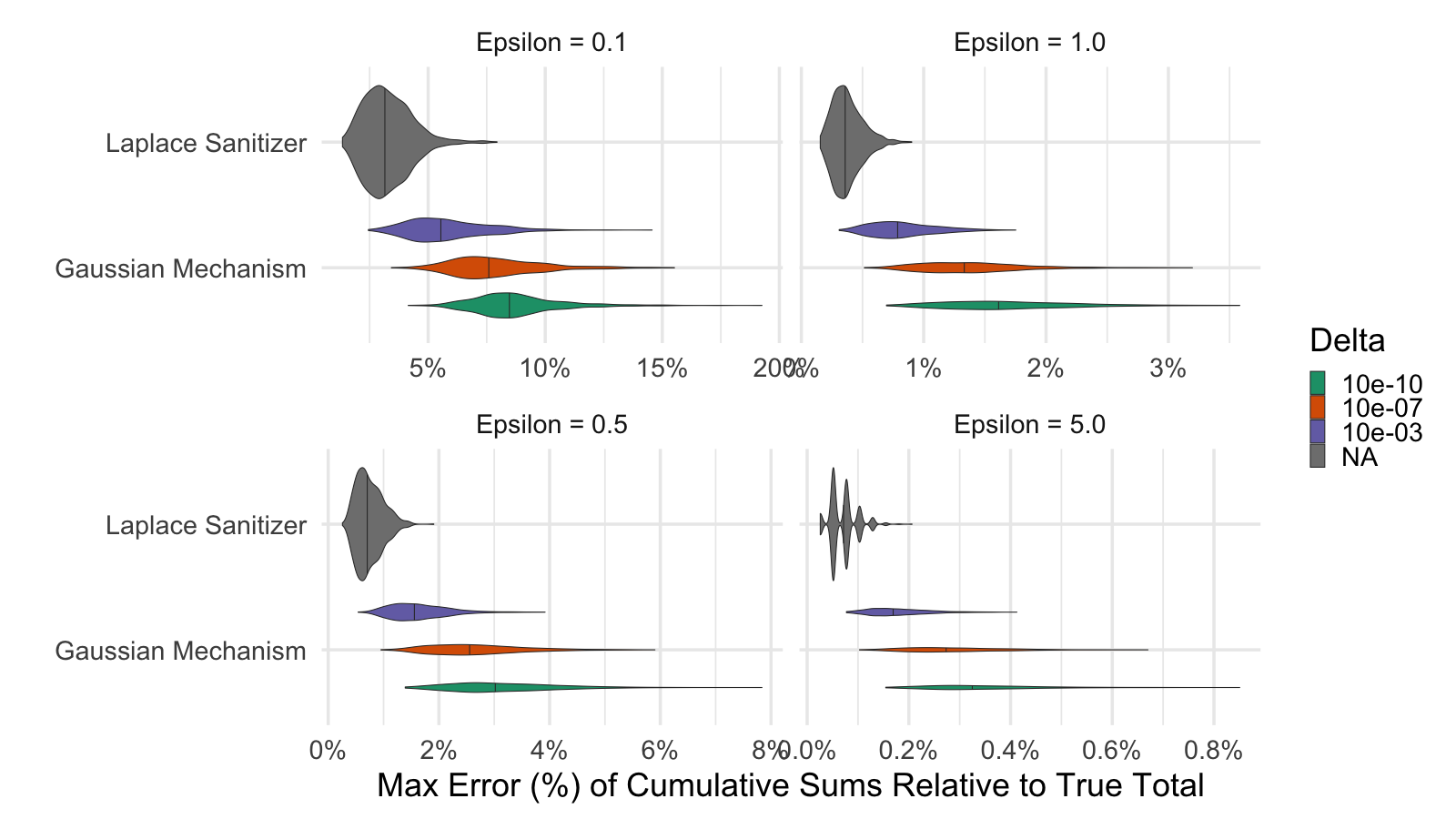}
    \caption{SOI Income histogram results: max relative error for two methods and varying levels of $\epsilon$ and $\delta$.}
    \label{fig:soi_inc_hist_max_supplemental}
\end{figure}

\begin{figure}[!htb]
    \centering
    \includegraphics[width=\textwidth]{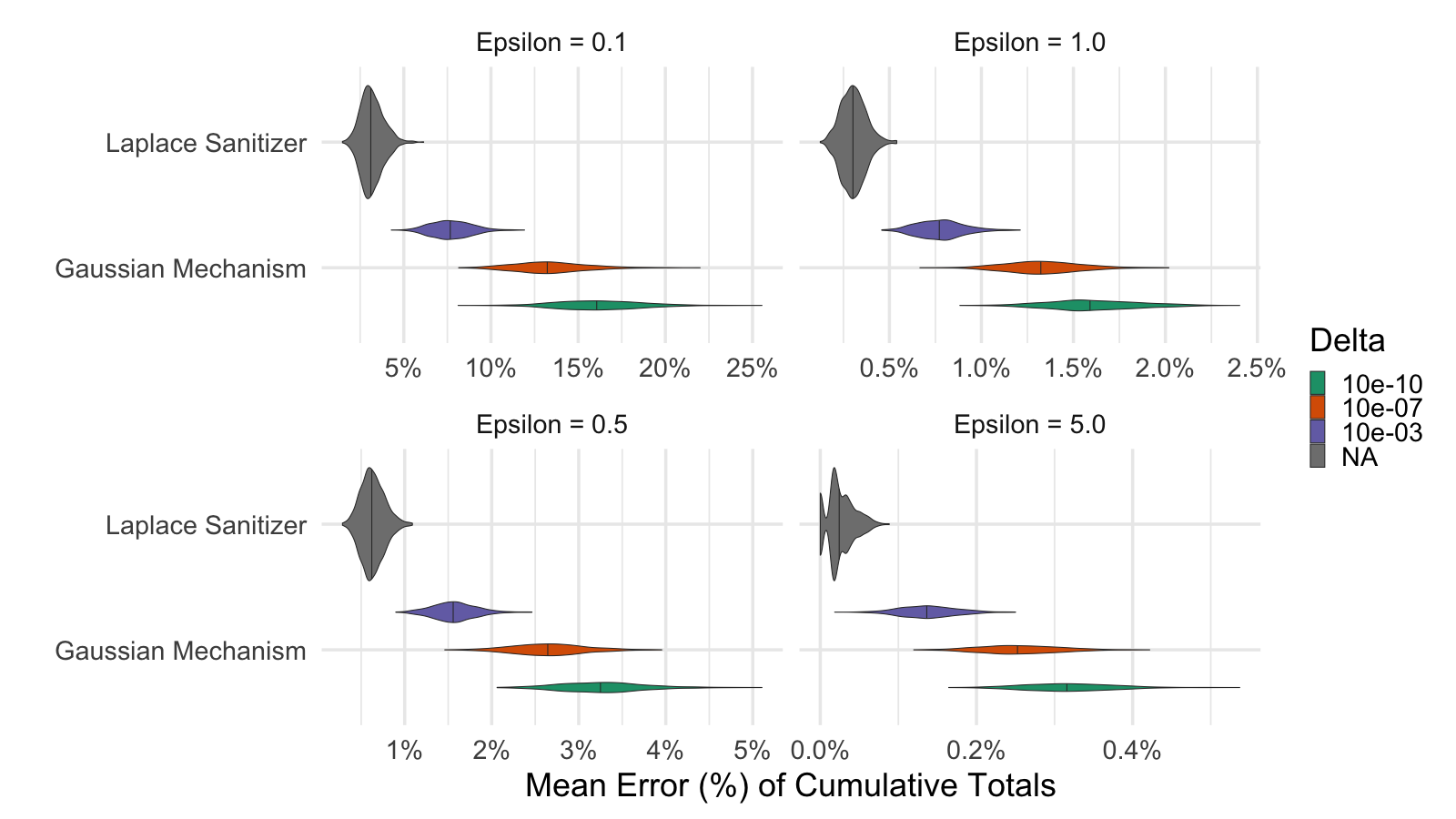}
    \caption{SOI Income histogram results: mean cumulative sums error for two methods and varying levels of $\epsilon$ and $\delta$.}
    \label{fig:soi_inc_hist_mean_supplemental}
\end{figure}

\newpage \clearpage
\subsection{Quantiles}
\begin{table}[!htp]
    \def\arraystretch{1}
    \caption{Summary of the SOI quantile results. Median simulated values for the mean errors across 9 quantiles are provided and 99th percentile simulated mean errors are in parenthesis. Select values of $\epsilon$ provided. For \textit{AppIndExp} and \textit{Smooth}, $\delta = 10^{-7}$}.
    \label{tab:soi_quant_results_supplemental}
    \centering
    \scriptsize
    \small
    \begin{tabular}{ C{3cm} | C{2cm} | C{4cm} | C{5.5cm}}
        \hline
            \textbf{Method} & \textbf{Epsilon} & \textbf{Mean Absolute Bias (\$)} & \textbf{Mean Ordered Differences Absolute Bias (\$)} \\
        \hline
            \textit{AppIndExp} & 0.01 & 6,867 (2,822,716) & 5,869 (2,826,831) \\
            \textit{JointExp} & 0.01 & 49,865 (1,732,398) & 26,753 (1,691,948) \\
            \textit{Smooth} & 0.01 & 23,877,189 (41,592,581) & 5,897,289 (5,897,289) \\
            \hline
            \textit{AppIndExp} & 0.1 & 579 (2,659) & 451 (3,012) \\
            \textit{JointExp} & 0.1 & 384 (938) & 244 (634) \\
            \textit{Smooth} & 0.1 & 278,636 (5,604,680) & 213,595 (5,460,180) \\
            \hline
            \textit{AppIndExp} & 1 & 50 (304) & 34 (238) \\
            \textit{JointExp} & 1 & 34 (74) & 11 (51) \\
            \textit{Smooth} & 1 & 6,525 (46,023) & 6,602 (44,974) \\
            \hline
            \textit{AppIndExp} & 5 & 31 (95) & 16 (63) \\
            \textit{JointExp} & 5 & 35 (44) & 5 (9) \\
            \textit{Smooth} & 5 & 316 (1,179) & 322 (1,379) \\
            \hline
        \hline
    \end{tabular}
\end{table}

\begin{figure}[!htb]
    \centering
    \includegraphics[width=\textwidth]{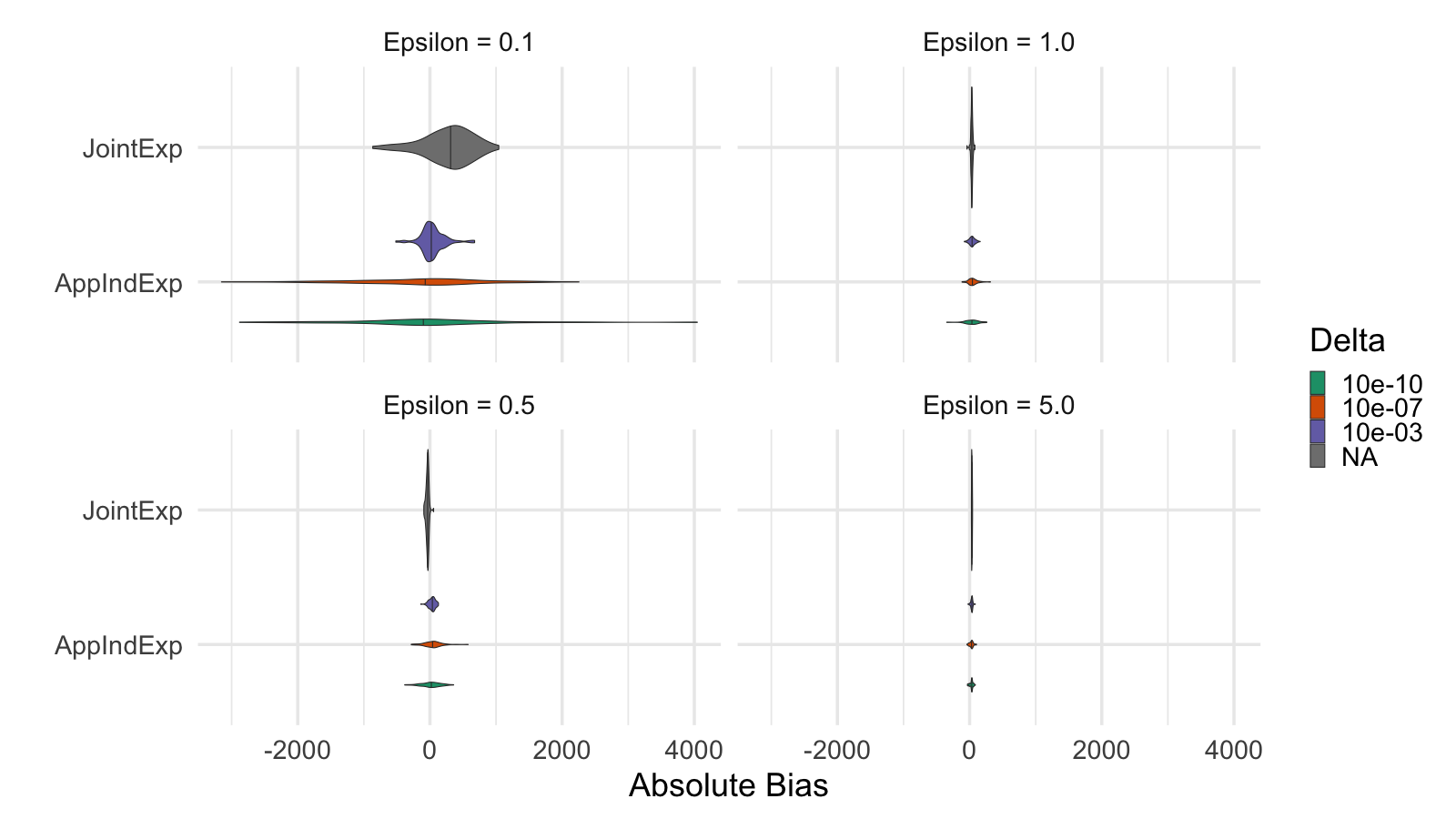}
    \caption{Average absolute bias across SOI quantiles for two methods and varying levels of $\epsilon$ and $\delta$.}
    \label{fig:soi_quantile_abs_supplemental}
\end{figure}

\newpage \clearpage
\subsection{Estimates of the Mean and Confidence Intervals}
\begin{table}[!htp]
    \def\arraystretch{1}
    \caption{Summary of the SOI confidence interval measures for the mean statistic results. Median simulated values for the metrics are provided and 90th percentile (worse) simulated values are in parenthesis. Select values of $\epsilon$ provided. For BHM, $\delta = 10^{-7}$.}
    \label{tab:soi_mean_ci_results_supplemental}
    \centering
    \scriptsize
    \small
    \begin{tabular}{ C{4cm} | C{2cm} | C{3.25cm} | C{3cm}}
        \hline
            \textbf{Method} & \textbf{Epsilon} & \textbf{CIR} & \textbf{CIO} \\
        \hline
            \textit{NOISYMAD} & 0.1 & 2.24 (2.40) & 0.72 (0.69) \\
            \textit{NOISYVAR} & 0.1 & 2.52 (2.75) & 0.69 (0.61) \\
            \textit{BHM} & 0.1 & 13.5 (19.7) & 0.54 (0.53) \\
            \hline
            \textit{NOISYMAD} & 0.5 & 0.58 (0.61) & 0.79 (0.77) \\
            \textit{NOISYVAR} & 0.5 & 1.09 (1.22) & 0.93 (0.82) \\
            \textit{BHM} & 0.5 & 3.33 (4.74) & 0.65 (0.60) \\
            \hline
            \textit{NOISYMAD} & 1 & 0.45 (0.46) & 0.72 (0.71) \\
            \textit{NOISYVAR} & 1 & 1.02 (1.09) & 0.97 (0.91) \\
            \textit{BHM} & 1 & 1.92 (2.58) & 0.75 (0.69) \\
            \hline
            \textit{NOISYMAD} & 5 & 0.40 (0.42) & 0.70 (0.69) \\
            \textit{NOISYVAR} & 5 & 1.00 (1.03) & 0.99 (0.97) \\
            \textit{BHM} & 5 & 1.04 (1.34) & 0.90 (0.83) \\
        \hline
    \end{tabular}
\end{table}

\begin{figure}[!htb]
    \centering
    \includegraphics[width=\textwidth]{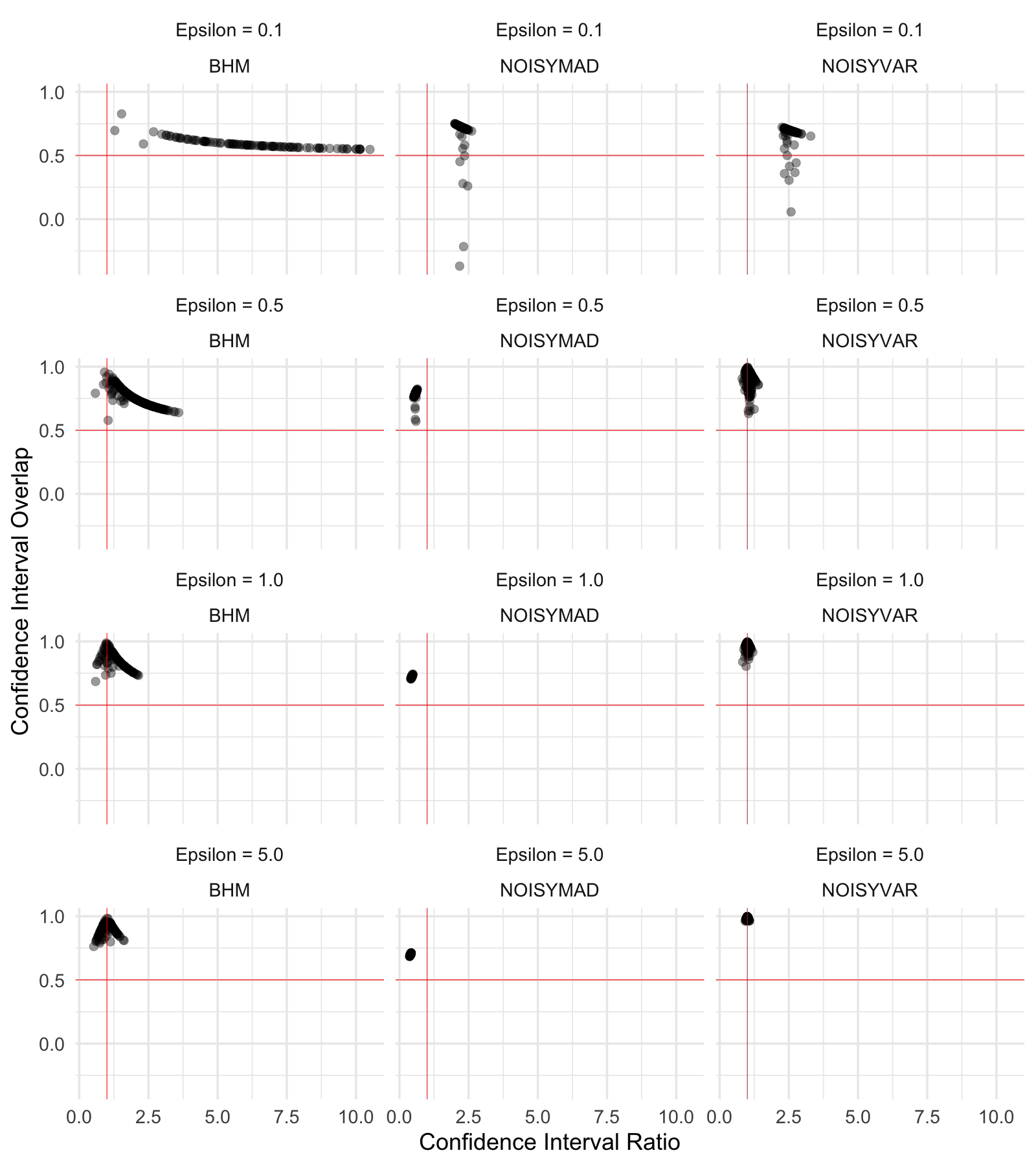}
    \caption{Mean SOI income confidence intervals ratio and overlap results for three methods and varying levels of $\epsilon$. BHM Results Only Shown for $\delta = 0.01$.}
    \label{fig:soi_mean_inc_ci_supplemental}
\end{figure}

\newpage \clearpage
\subsection{Regression Models}
\begin{figure}[!htb]
    \centering
    \includegraphics[width=6in]{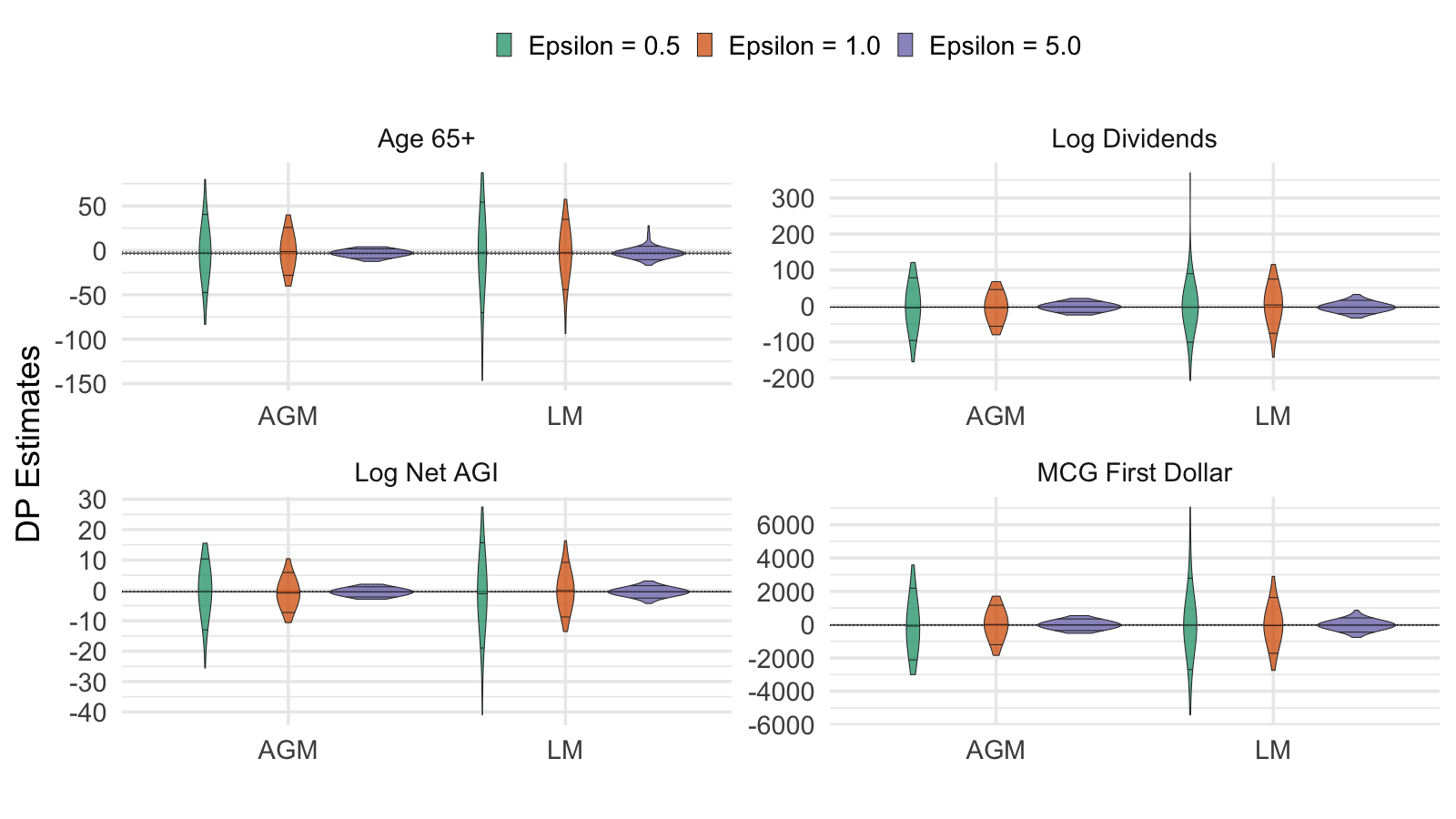}
    \caption{Distribution of simulated DP estimates for SOI regression coefficients. Confidential estimates (black horizontal lines) and 95\% confidence intervals (dotted lines) are shown. $\delta = 10^{-7}$ for AGM.}
    \label{fig:soi_reg_abs_bias_all}
\end{figure}

\begin{table}[!htp]
    \def\arraystretch{1}
    \caption{Summary of the SOI regression results. Confidential (non-DP) estimate values for each coefficient are shown. 5th, 95th, quantiles, and L1 Error for select coefficients, methods, and values of $\epsilon$. $\delta = 10^{-7}$ for AGM.}
    \label{tab:soi_reg_results_error}
    \centering
    \scriptsize
    \small
    \begin{tabular}{C{4.75cm} | C{1.5cm} | C{1.6cm} | C{1.75cm} | C{1.75cm} | C{1.5cm}}
        \hline
            \textbf{Coefficient (estimate)} & \textbf{Epsilon} & \textbf{Method} & \textbf{DP 5\%} & \textbf{DP 95\%} & \textbf{$l_1$} \\
            \hline
            \multirow{ 6}{*}{Age 65+ (-3.16)} & 0.5 & \textit{AGM} & -51.8 & 36.8 &  20.3 \\
            & 0.5 & \textit{LM} & -91.8 & 66.4 & 31.5 \\\cline{2-6}
            & 1 & \textit{AGM} & -29.8 & 27.4 & 14.1 \\
            & 1 & \textit{LM} & -40.0 & 34.2 & 19.4 \\\cline{2-6}
            & 5 & \textit{AGM} & -8.49 & 2.78 & 2.66 \\
            & 5 & \textit{LM} & -10.5 & 5.01 & 3.9 \\
            \hline
            \hline
            \multirow{ 6}{*}{Log Dividends (-3.53)} & 0.5 & \textit{AGM} & -82.8 & 77.6 & 42.1 \\
            & 0.5 & \textit{LM} & -103 & 84.3 & 45.0 \\\cline{2-6}
            & 1 & \textit{AGM} & -58.1 & 45.0 & 27.4 \\
            & 1 & \textit{LM} & -70.1 & 67.9 & 35.4 \\\cline{2-6}
            & 5 & \textit{AGM} & -19.0 & 10.4 & 7.80 \\
            & 5 & \textit{LM} & -21.8 & 17.7 & 8.68 \\
            \hline
            \hline
            \multirow{ 6}{*}{Log Net AGI (-0.433)} & 0.5 & \textit{AGM} & -12.7 & 10.6 & 5.57 \\
            & 0.5 & \textit{LM} & -19.6 & 16.4 & 7.96 \\\cline{2-6}
            & 1 & \textit{AGM} & -6.42 & 5.18 & 3.27 \\
            & 1 & \textit{LM} & -9.68 & 9.60 & 4.24 \\\cline{2-6}
            & 5 & \textit{AGM} & -2.34 & 1.26 & 0.878 \\
            & 5 & \textit{LM} & -2.35 & 1.70 & 0.943 \\
            \hline
            \hline
            \multirow{ 6}{*}{MCG First Dollar (-28.4)} & 0.5 & \textit{AGM} & -2170 & 2290 & 1180 \\
            & 0.5 & \textit{LM} & -3130 & 3860 & 1340 \\\cline{2-6}
            & 1 & \textit{AGM} & -1370 & 1130 & 598 \\
            & 1 & \textit{LM} & -1810 & 1710 & 840 \\\cline{2-6}
            & 5 & \textit{AGM} & -346 & 343 & 175 \\
            & 5 & \textit{LM} & -479 & 368 & 196 \\
            \hline
    \end{tabular}
\end{table}

\begin{table}[!htp]
    \def\arraystretch{1}
    \caption{Summary of the SOI private CI estimates. Median and 90th Percentile Values CIR and Percentage of Sign, Significance Agreement, and CI Overlap for select coefficients, methods, and values of $\epsilon$. Both methods for estimating CI shown. $\delta = 10^{-7}$ for AGM.}
    \label{tab:soi_reg_results_ci}
    \centering
    \scriptsize
    \small
    \begin{tabular}{ C{3.5cm} | C{1.5cm} | C{1.6cm}| C{2cm} | C{1.25cm} | C{2cm} | C{1.25cm}}
            \hline
            \multicolumn{3}{c|}{} & \multicolumn{2}{|c|}{\textbf{Asymptotic}} & \multicolumn{2}{|c}{\textbf{Bootstrap}} \\
            \hline
            \textbf{Coefficient} & \textbf{Epsilon} & \textbf{Method} & \textbf{CIR} & \textbf{SSO\%} & \textbf{CIR} & \textbf{SSO\%} \\
            \hline
            \multirow{ 6}{*}{Age 65+} & 0.5 & \textit{AGM} & 2.79 (3.95) & 9\% & 39.6 (44.8) & 0\% \\
            & 0.5 & \textit{LM} & 4.02 (5.38) & 4\% & 64.6 (74.0) & 1\% \\\cline{2-7}
            & 1 & \textit{AGM} & 2.00 (2.08) & 9\% & 20.6 (21.3) & 3\% \\
            & 1 & \textit{LM} & 3.13 (3.82) & 4\% & 32.4 (35.3) & 0\% \\\cline{2-7}
            & 5 & \textit{AGM} & 1.01 (1.39) & 45\%  & 4.88 (5.00) & 11\% \\
            & 5 & \textit{LM} & 1.17 (1.65) & 46\%  & 6.58 (6.73) & 7\% \\
            \hline
            \hline
            \multirow{ 6}{*}{Log Dividends} & 0.5 & \textit{AGM} & 2.59 (3.36) & 2\% & 210 (359) & 0\% \\
             & 0.5 & \textit{LM} & 3.10 (4.26) & 2\% & 244 (540) & 0\% \\\cline{2-7}
             & 1 & \textit{AGM} & 1.96 (2.47) & 5\% & 124 (161) & 0\% \\
             & 1 & \textit{LM} & 2.65 (3.38) & 2\% & 172 (267) & 2\% \\\cline{2-7}
             & 5 & \textit{AGM} & 1.02 (1.38) & 10\% & 33.3 (34.6) & 5\% \\
             & 5 & \textit{LM} & 1.16 (1.60) & 9\% & 45.3 (47.9) & 1\% \\
            \hline
            \hline
            \multirow{ 6}{*}{Log Net AGI} & 0.5 & \textit{AGM} & 2.68 (3.80) & 1\% & 120 (182) & 0\% \\
             & 0.5 & \textit{LM} & 3.85 (5.03) & 1\% & 163 (310) & 0\% \\\cline{2-7}
             & 1 & \textit{AGM} & 1.98 (2.73) & 7\% & 67.5 (80.8) & 4\% \\
             & 1 & \textit{LM} & 3.07 (3.68) & 3\% & 100 (132) & 2\% \\\cline{2-7}
             & 5 & \textit{AGM} & 1.01 (1.37) & 17\%  & 16.6 (17.0) & 8\% \\
             & 5 & \textit{LM} & 1.15 (1.63) & 24\%  & 22.2 (23.1) & 0\% \\
            \hline
            \hline
            \multirow{ 6}{*}{MCG First Dollar} & 0.5 & \textit{AGM} & 2.53 (3.48) & 3\% & 164 (290) & 0\% \\
             & 0.5 & \textit{LM} & 3.37 (4.44) & 2\% & 214 (431) & 0\% \\\cline{2-7}
             & 1 & \textit{AGM} & 1.93 (2.53) & 3\% & 99.5 (123) & 0\% \\
             & 1 & \textit{LM} & 2.73 (3.49) & 1\% & 145 (208) & 0\% \\\cline{2-7}
             & 5 & \textit{AGM} & 1.01 (1.38) & 7\% & 25.1 (26.1) & 3\% \\
             & 5 & \textit{LM} & 1.14 (1.63) & 11\% & 34.5 (35.9) & 0\% \\
            \hline
    \end{tabular}
\end{table}

\begin{figure}[!htb]
    \centering
    \includegraphics[width=6.5in]{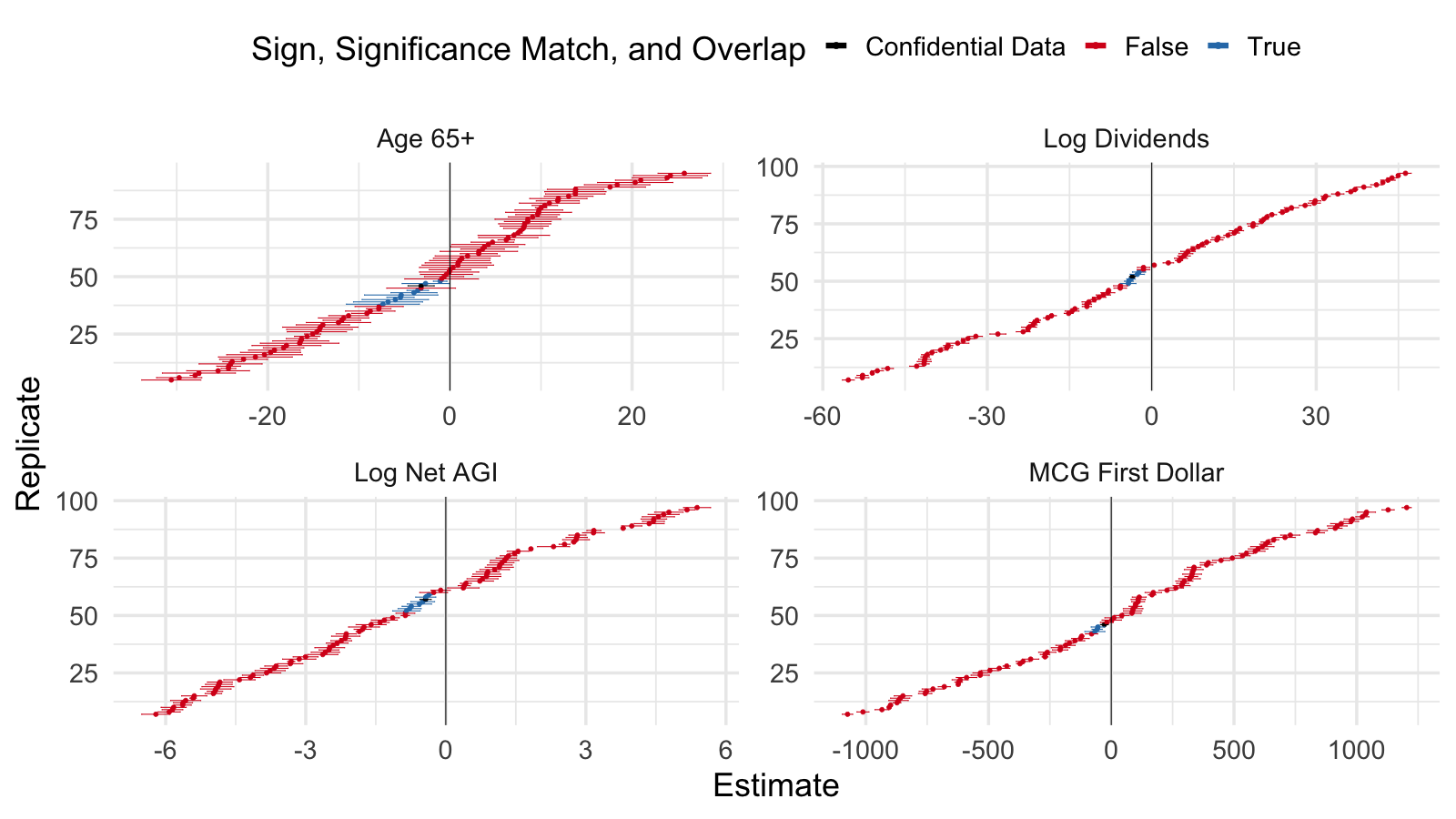}
    \caption{SOI regression results showing the 90 best (out of 100) simulated DP estimates and confidence intervals for the Analytic Gaussian Mechanism and $\epsilon = 1$. Results shown for two coefficients with confidence intervals estimated using the asymptotic approach. $\delta = 10^{-7}$ for AGM.}
    \label{fig:soi_reg_catepillar_asymptotic}
\end{figure}

\begin{figure}[!htb]
    \centering
    \includegraphics[width=6.5in]{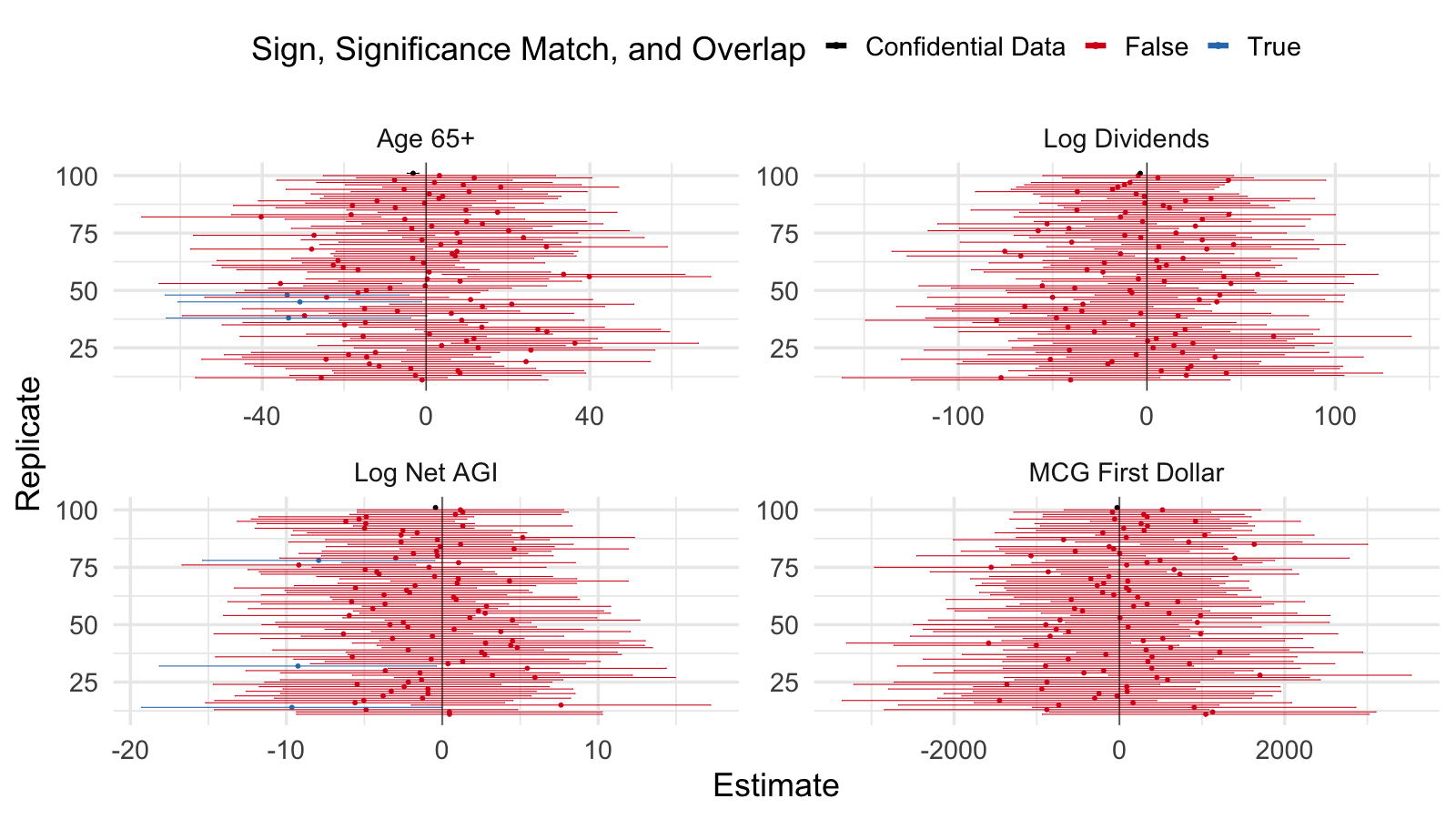}
    \caption{SOI regression results showing the 90 best (out of 100) simulated DP estimates and confidence intervals for the Analytic Gaussian Mechanism and $\epsilon = 1$. Results shown for two coefficients with confidence intervals estimated using the bootstrap approach. $\delta = 10^{-7}$ for AGM.}
    \label{fig:soi_reg_catepillar_boot}
\end{figure}

\newpage\clearpage
\bibliographystyle{chicago}
\bibliography{ref}

\end{document}